\documentclass[aps,prb,amsmath,twocolumn,amssymb,floatfixng,showpacs,
superscriptaddress,footinbib,multicol]{revtex4-1}
\usepackage{graphicx}
\usepackage{epstopdf}
\usepackage{dcolumn}
\usepackage{bm}
\usepackage{color}
\usepackage{amsmath}
\usepackage{braket}
\usepackage{amsfonts}
\usepackage{soul}
\usepackage{url}
\usepackage{array,tabularx}
\usepackage{graphics}
\usepackage{times}
\usepackage{comment}
\usepackage{subfigure}
\usepackage[export]{adjustbox}
\usepackage[colorlinks=true,linktoc=page,linkcolor=red,citecolor=blue]{hyperref}
\newcommand\x{4.8}
\newcommand\y{3.7}

\begin{document}

\title{Tunneling-like wave transmission in non-Hermitian lattices with mirrored nonreciprocity}
\author{Sayan Jana}\email{sayanjana@tauex.tau.ac.il}
\affiliation{School of Mechanical Engineering, Tel Aviv University, Tel Aviv 69978, Israel}
\author{Lea Sirota}\email{leabeilkin@tauex.tau.ac.il}
\affiliation{School of Mechanical Engineering, Tel Aviv University, Tel Aviv 69978, Israel}

\begin{abstract}

We report a peculiar tunneling phenomenon that occurs in lattices with nonreciprocal couplings.
The nonreciprocity holds for an inner portion of the lattice, constituting a non-Hermitian interface between outer Hermitian sections.
The couplings are mirrored about the interface center.
As a standalone system that was widely studied in recent years, each section of the interface supports the non-Hermitian skin effect, in which modes are accumulated at one boundary. 
Here, we investigate what happens to a wave that propagates along the lattice and hits the interface.
The skin mode accumulation, which effectively constitutes a barrier, forbids wave penetration into the interface, but surprisingly, under certain conditions the wave is transmitted to the other side, keeping the interface dark, as if the wave invisibly tunneled through it. 
Remarkably, the tunneling is independent of the interface length, and a perfect transmission can be achieved independently of frequency and nonreciprocity strength.
We derive the phenomenon both for quantum and classical systems, and realize it experimentally in an active topoelectric metamaterial. 
Our study fosters the research of wave tunneling through other types of non-Hermitian interfaces, which may also include nonlinearities, time-dependence and more. 

\end{abstract}

\maketitle

\begin{figure}[htbp]
\begin{center}
\begin{tabular}[t]{c}
    \begin{tabular}[t]{c}
\textbf{(a)} \\
      \includegraphics[width=8.2 cm]{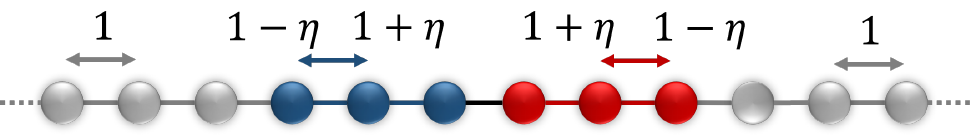} 
\end{tabular} \\
\setlength{\tabcolsep}{-3pt}
\def\arraystretch{0.9} 
    \begin{tabular}[t]{c c}
     \textbf{(b)} & \textbf{(c)} \\
\includegraphics[height=3.2 cm, valign=c]{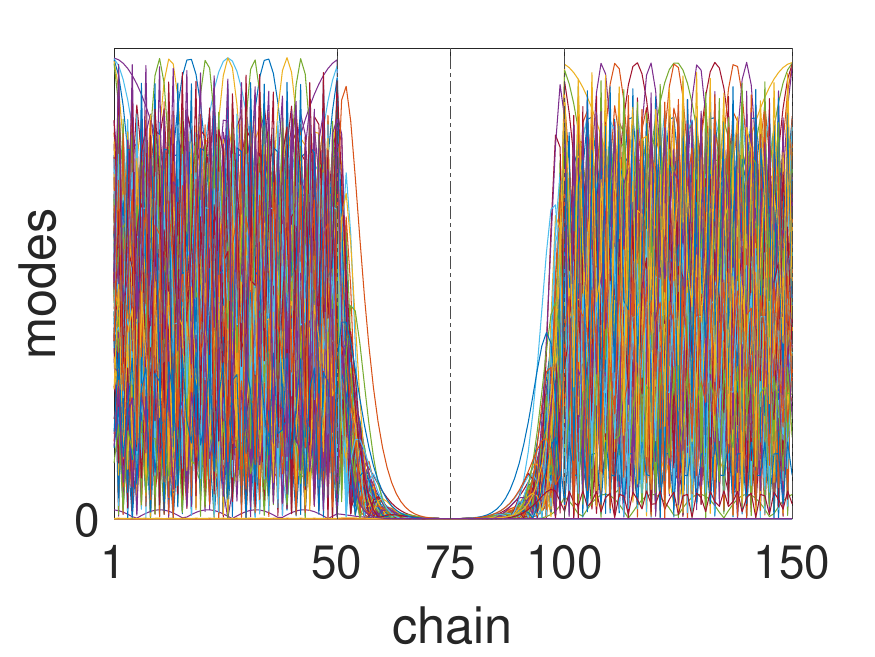} &
\includegraphics[height=3.3 cm, valign=c]{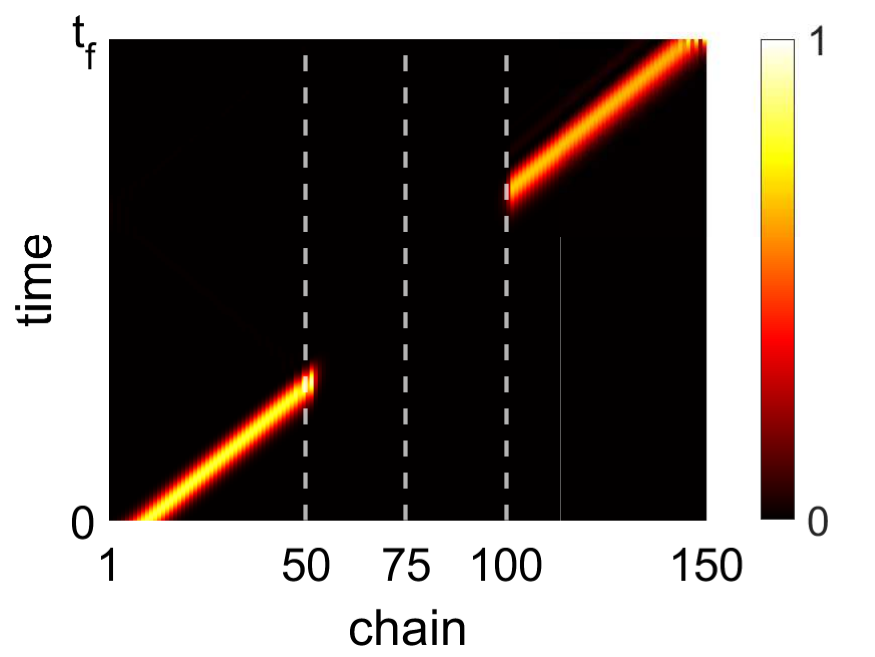} \\
\textbf{(d)} & \textbf{(e)} \\
      \includegraphics[height=3.2 cm, valign=c]{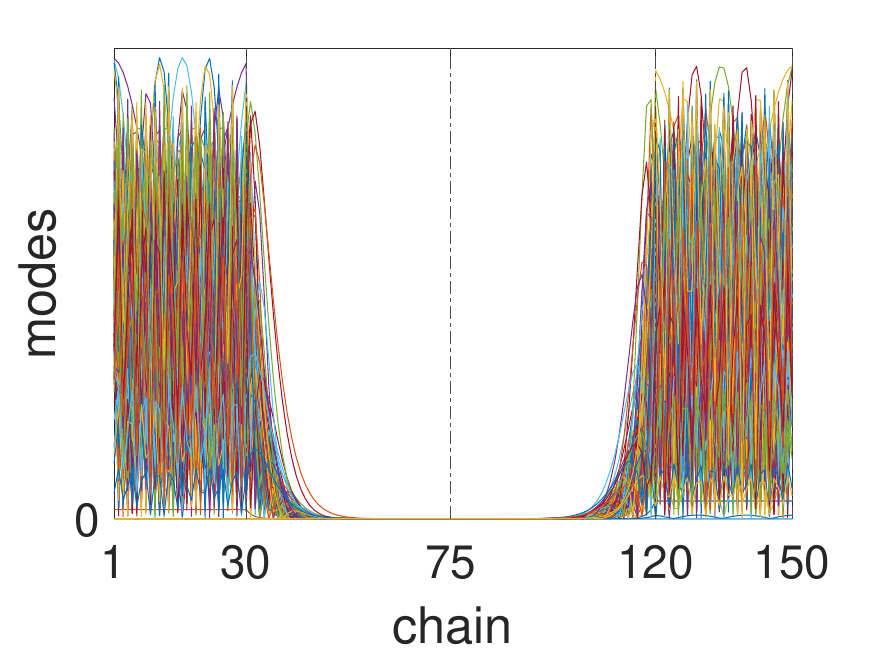} & \includegraphics[height=3.3 cm, valign=c]{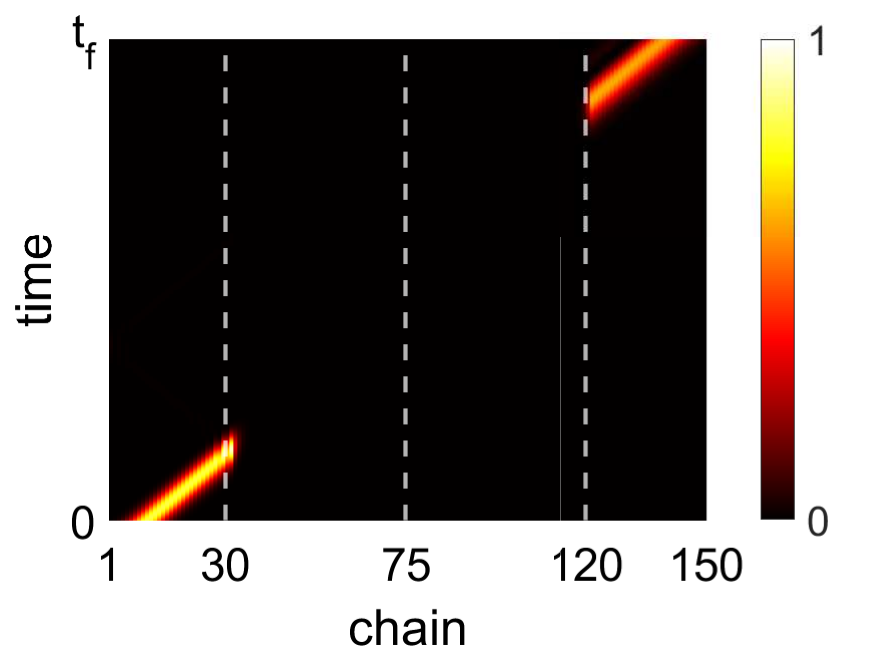} 
\end{tabular}
\end{tabular}
\end{center}
\caption{\textbf{The quantum model.} (a) Lattice schematic partitioned into two Hermitian sections (gray) and two non-Hermitian sections (blue, red). 
(b)-(c) Eigenmodes and time domain numerical simulation of a 150 sites lattice with $50$ non-Hermitian sites, and $\eta$=0.5, demonstrating wavepacket tunneling-like transmission through the interface. (d)-(e) The same as in (b)-(c) for $90$ non-Hermitian sites.}
\label{fig:model}
\end{figure}

Non-Hermitian physics has received significant attention in recent years, both for quantum and classical systems. 
This physics refers to non-conservative systems that interact with the environment\cite{berry2004physics}, 
where commonly, the interaction is obtained 
by the addition of gain and loss,
or by inducing 
some form of nonreciprocity in the system. 
The resulting spectrum, usually complex-valued, has shed new light on the originally Hermitian concepts of topological invariants, bulk-boundary correspondence and its breakdown, and the associated topological protection of boundary modes \cite{yao2018edge,ashida2020non}.
The underlying wave dynamics unveiled new exotic regimes of wave propagation in photonic, acoustic, and elastic systems, such as 
unidirectional invisibility, cloaking, coherent absorption, and more \cite{makris2008beam,zhu2014p,sounas2015unidirectional,fleury2015invisible,shi2016accessing,achilleos2017non,gu2021acoustic,cao2022design,el2018non,gu2021controlling,huang2023acoustic}. 

For non-Hermiticity due to non-reciprocity, which can be induced both between local and long distance sites of a lattice, the wave propagation is usually amplified in one direction and attenuated in the other. 
This inherent unidirectional dynamics is related to the celebrated skin effect
\cite{lee2019anatomy,goldsberry2019non,nassar2020nonreciprocity,scheibner2020non,rosa2020dynamics,helbig2020generalized,rasmussen2021acoustic,zhang2021acoustic}. 
The extraordinary underlying topology 
gives rise to accumulation of the skin modes at the lattice boundary, where the attracting boundary can be switched by flipping the polarization of the non-reciprocity parameters.
While the skin effect is obtained for nonreciprocity along an entire non-Hermitian lattice, or a chain in one dimension, it turns out, as we report in this work, that intriguing wave dynamics can be supported at the intersection of Hermitian and non-Hermitian chains.

In particular, a tunneling-like effect emerges when connecting two non-Hermitian nonreciprocal chains, as depicted by blue and red in the schematic of Fig. \ref{fig:model}(a), placed as an interface between Hermitian chains, depicted by gray. 
Known tunneling phenomena, which originate from the quantum realm and sometimes are emulated in classical systems, targets Hermitian barriers. This includes, for example, the Klein tunneling of relativistic particles
through barriers of arbitrary heights and widths \cite{katsnelson2006chiral,jiang2020direct,sirota2022klein,zhu2023experimental}, or tunneling of particles across the event horizon of black holes
\cite{volovik2016black,liang2019curved,kedem2020black,sabsovich2022hawking,jana2023gravitational}. 
Here, we derive and experimentally demonstrate a tunneling-like phenomenon of waves through a non-Hermitian barrier that features a nonreciprocal structural force, and study the fundamental differences of this type of tunneling from its Hermitian counterparts.

We refer again to the lattice schematic in Fig. \ref{fig:model}(a).
For the gray sections we set a normalized coupling of unity between neighboring sites, equal in both directions. In contrast, the blue section has the electron creation operator at each site coupled to the annihilation operator of its nearest neighbor to the right with a stronger coupling of 1+$\eta$, and to the left with a weaker coupling of 1-$\eta$, where $\eta\in (0,1)$. For the red section this definition is mirrored. The nonreciprocal sections follow the Hatano-Nelson model, and are governed by the tight-binding Hamiltonian
\begin{equation}  \label{eq:Hamiltonian}      H=\textstyle{\sum}_{j}\left(1+\eta\right)\alpha_{j}^{\dagger}\alpha_{j+1}+\left(1-\eta\right)\alpha_{j+1}^{\dagger}\alpha_{j},
\end{equation}
with $\eta>$0 for the blue and $\eta<$0 for the red, respectively.
Each standalone nonreciprocal chain with open boundaries supports the non-Hermitian skin effect, where skin modes are accumulated at the boundary of the 1-$\eta$ coupling direction \cite{guo2021exact,longhi2017nonadiabatic}. 
Therefore, for the assembly of the four sections of our model, which can be labeled as H1-NH1-NH2-H2 (H and NH stand for Hermitian and non-Hermitian, respectively), it would be expected for the modes to be supported along the Hermitian sections only, but not inside the non-Hermitian interface. This is indeed the case, as depicted in Figs. \ref{fig:model}(b) and (d) for two different interface sizes. 
Consequently, it would be expected that when a wavepacket propagating along a Hermitian section hits the interface, the energy will not pass it,
as it is not allowed beyond the boundary. 

However, it turns out that upon hitting the boundary, for the energy window $|E|<2\sqrt{1-\eta^2}$, the wave seemingly disappears,
and reemerges on the other side of the interface at a later time.
This unique dynamics portrays an effect as if the wave tunneled through the non-Hermitian interface, creating the associated dark region. 
The effect is demonstrated in Figs. \ref{fig:model}(c) and (e) by simulating the time domain evolution of a Gaussian wavepacket in a quantum lattice sectioned according to Fig. \ref{fig:model}(a), for two different interface sizes.
Notably, in both cases the transmission coefficient equals 0.986, implying that the tunneling effect is independent of the interface length\cite{supplementary}.

We now demonstrate that the phenomenon can be supported in classical systems. 
The blue and the red couplings of Fig. \ref{fig:model}(a) then indicate different compliance, $1-\eta$ and $1+\eta$, to energy transmission in each direction. This results in nonreciprocal propagation along the blue and red chains. The associated waves can be displacement waves in mechanical mass-spring lattices, voltage waves in electrical transmission lines, etc. 
%
We then derive a model analogous to \eqref{eq:Hamiltonian} for a lattice obeying second order dynamics of the form $\Ddot{\textbf{Y}}=D\textbf{Y}$. Here, $\textbf{Y}$ is the response vector and $D$ is the associated real-valued dynamical matrix
\cite{supplementary}. 
The $n_{th}$ rows of $D$ that correspond to the non-Hermitian sections are given by 
\begin{equation}  \label{eq:nonrecip_eq}
    \omega_0^{-2}\Ddot{Y}_n=Y_{n+1}-2Y_n+Y_{n-1}+\eta(Y_{n+1}-Y_{n-1}),
\end{equation}
where $\omega_0$ is the natural frequency.
The term $Y_{n+1}-Y_{n-1}$, which is equivalent to a first order spatial derivative, is associated with the non-Hermitian skin effect in classical systems \cite{brandenbourger2019non}. The underlying nonreciprocal wave grows in amplitude in the weaker coupling direction, 
and decays in the opposite direction, forcing the modes towards the boundaries. 
%
\begin{figure*}[htpb]
    \begin{center}
    \def\arraystretch{0.9} 
    \setlength{\tabcolsep}{1pt}
        \begin{tabular}{ccccc}
        \textbf{(a)} & \textbf{(b)} & \textbf{(c)} &  & \textbf{(d)} $\eta=0$ \\
           \includegraphics[height=3.5 cm, valign=c]{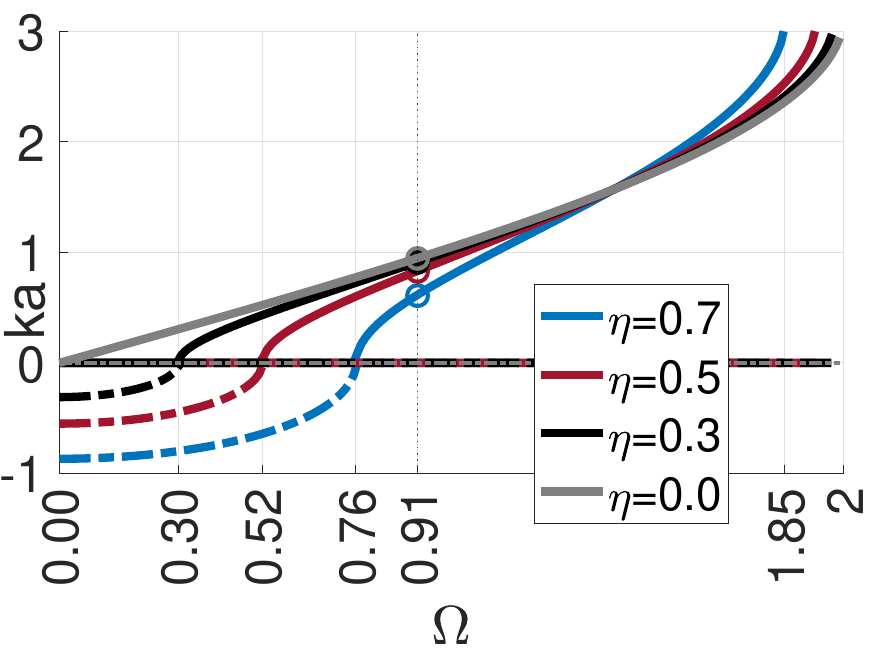} &  \includegraphics[height=3.5 cm, valign=c]{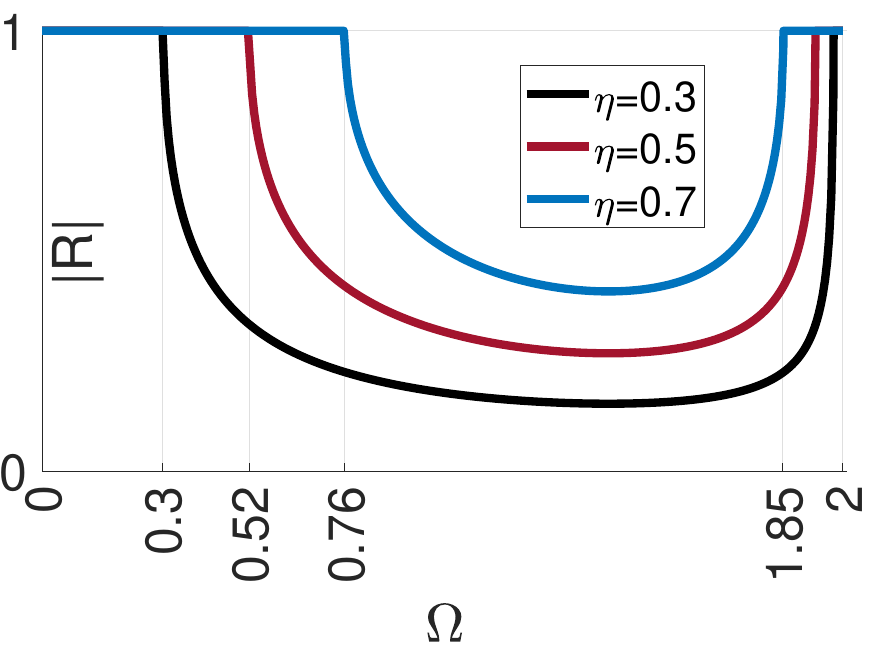}  &  \quad \includegraphics[height=3.5 cm, valign=c]{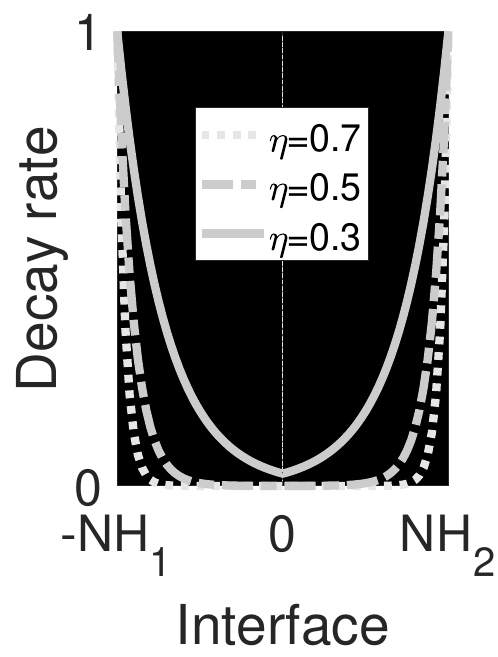} &  \: \includegraphics[height=2.6 cm, valign=c]{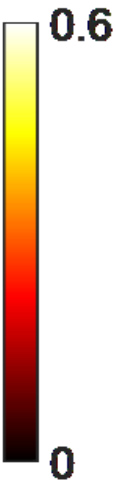} & \; \includegraphics[height=3.0 cm, valign=c]{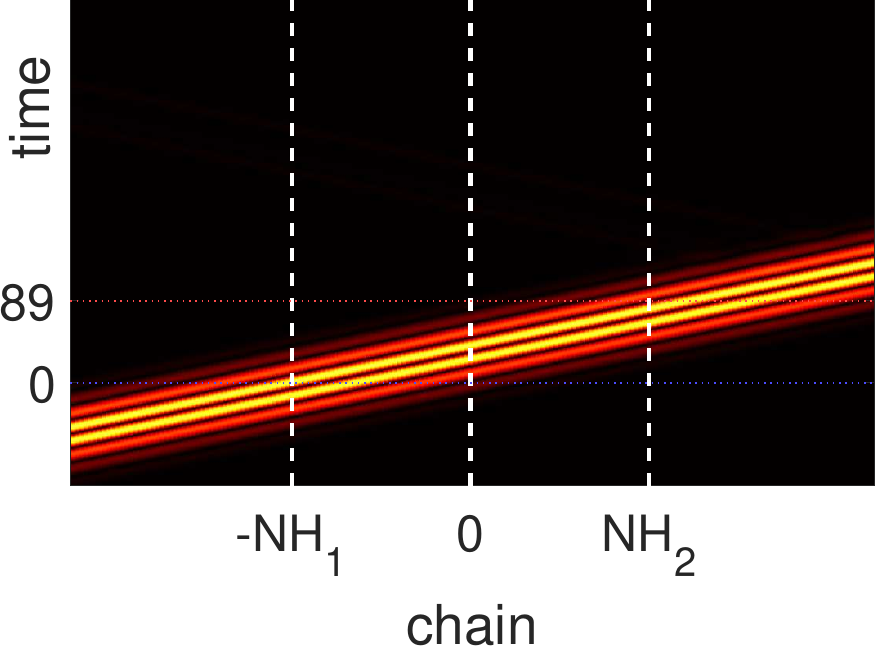} \end{tabular} \\ \setlength{\tabcolsep}{-1pt}
        \begin{tabular}{cccc}
            \textbf{(e)} $\eta=0.3$ & \textbf{(f)} $\eta=0.3$ & \textbf{(g)} $\eta=0.5$ & \textbf{(h)} $\eta=0.7$ \\
            \includegraphics[height=3.3 cm, valign=c]{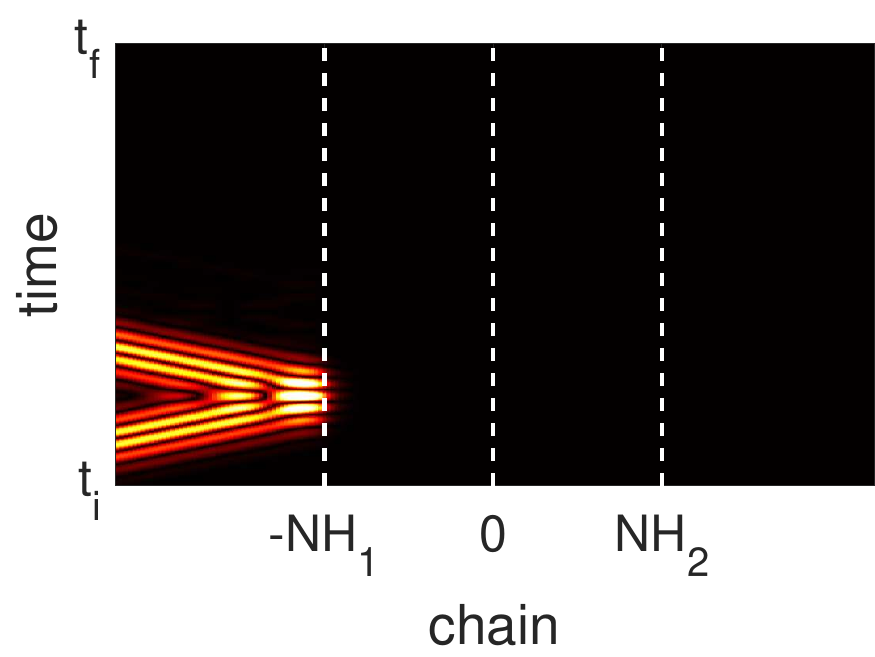} & \includegraphics[height=3.3 cm, valign=c]{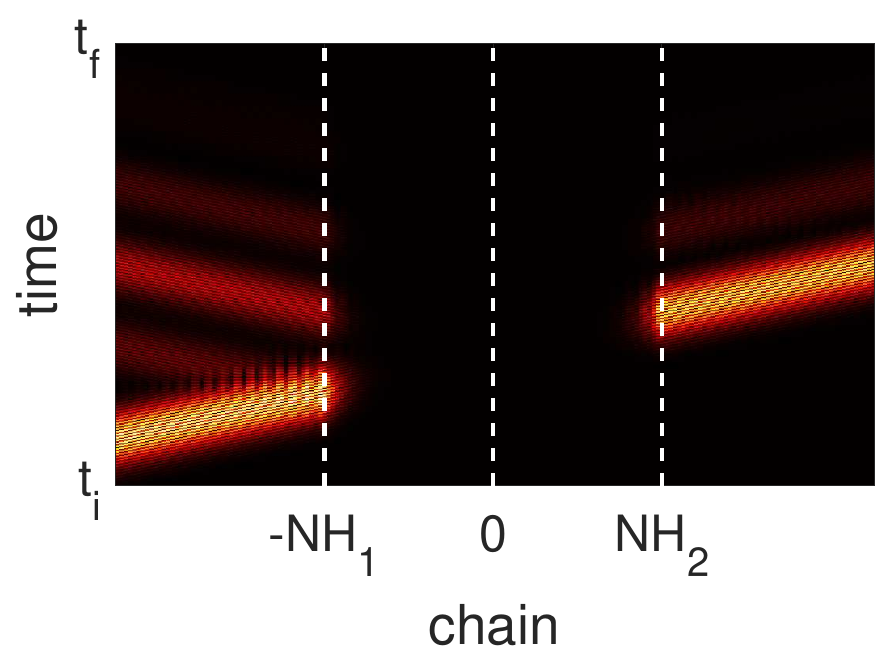} & \includegraphics[height=3.3 cm, valign=c]{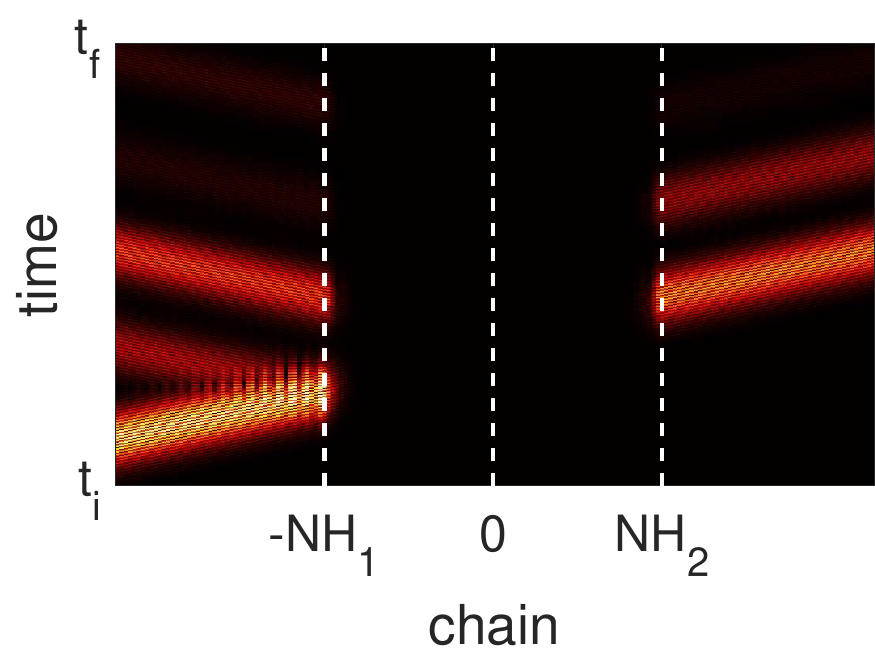} & \includegraphics[height=3.3 cm, valign=c]{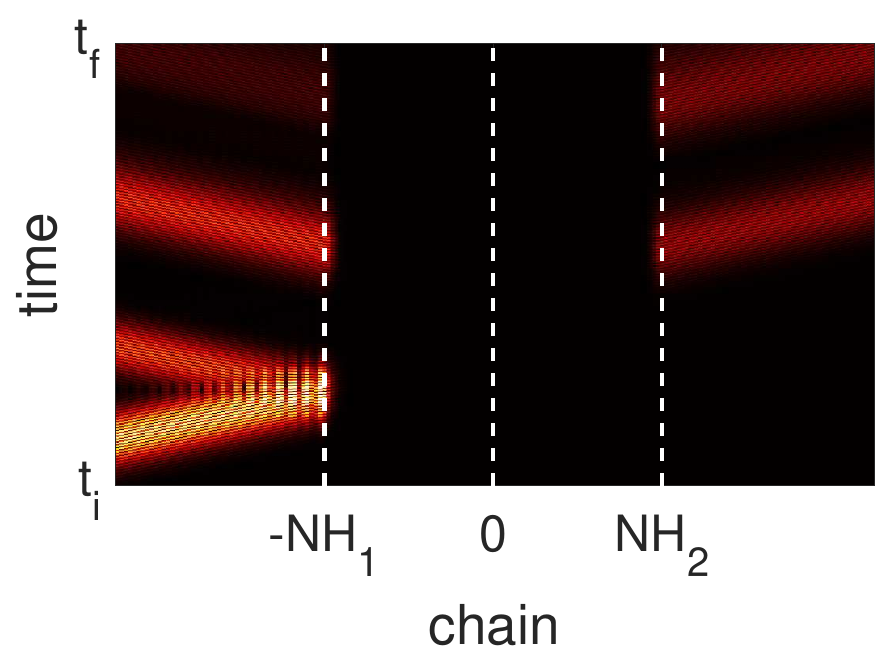}
             \\
            \textbf{(i)} $\eta=0.3$ & \textbf{(j)} $\eta=0.3$ & \textbf{(k)} $\eta=0.5$ & \textbf{(l)} $\eta=0.7$ \\
            \includegraphics[height=3.4 cm, valign=c]{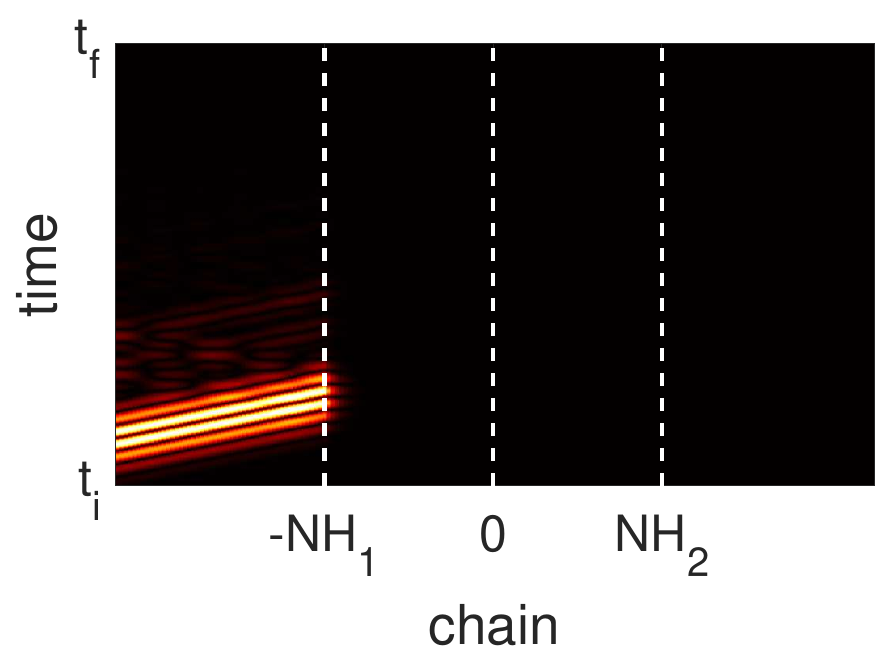} & \includegraphics[height=3.2 cm, valign=c]{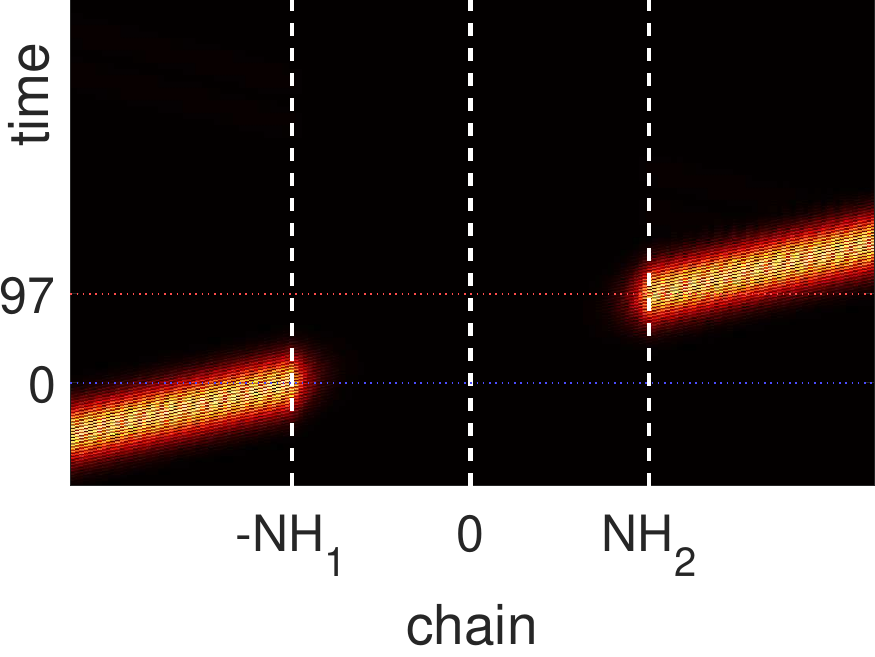} &  \includegraphics[height=3.2 cm, valign=c]{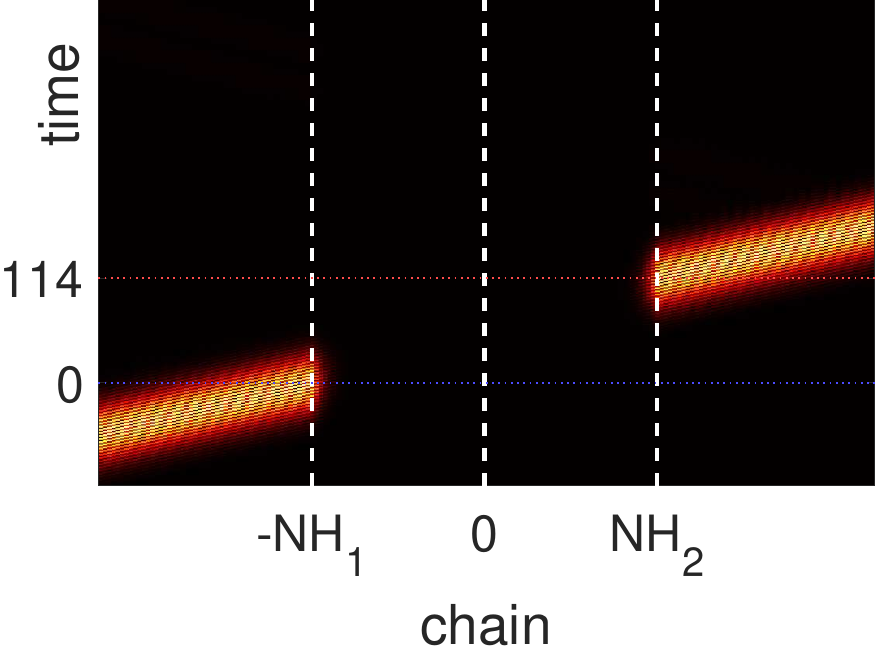} &  \includegraphics[height=3.2 cm, valign=c]{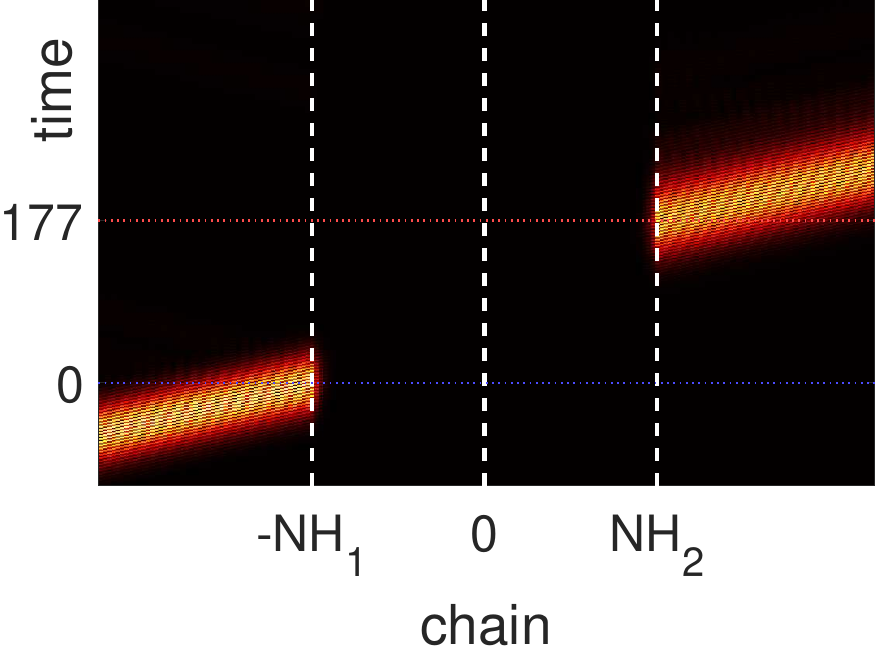}
            \\
            \textbf{(m)} & \textbf{(n)} $\eta=0.3$ & \textbf{(o)} $\eta=0.5$ & \textbf{(p)} $\eta=0.7$ \\
            \includegraphics[height=3.3 cm, valign=c]{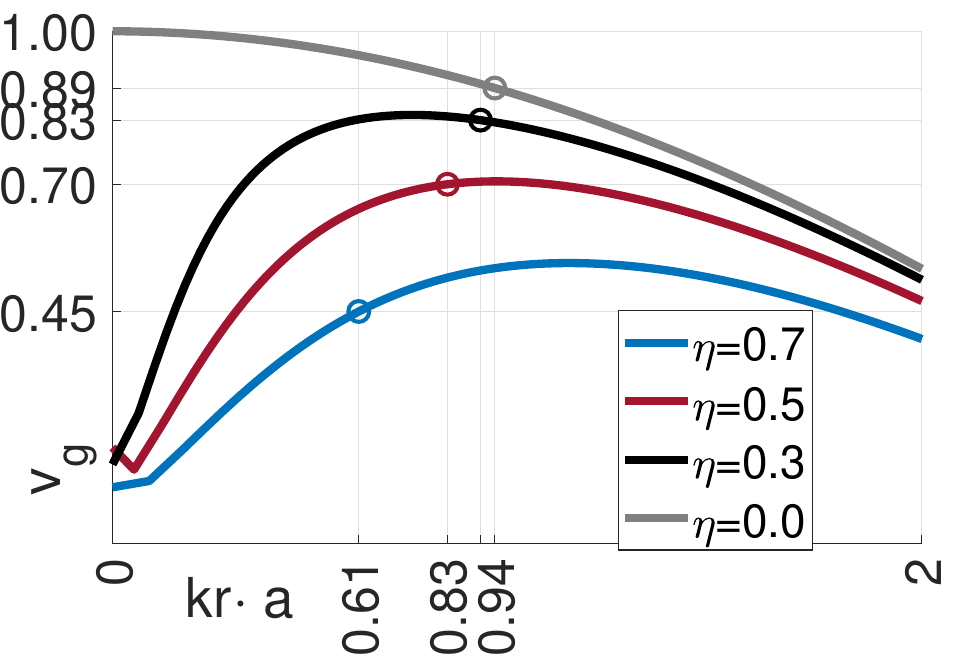} & \includegraphics[height=3.2 cm, valign=c]{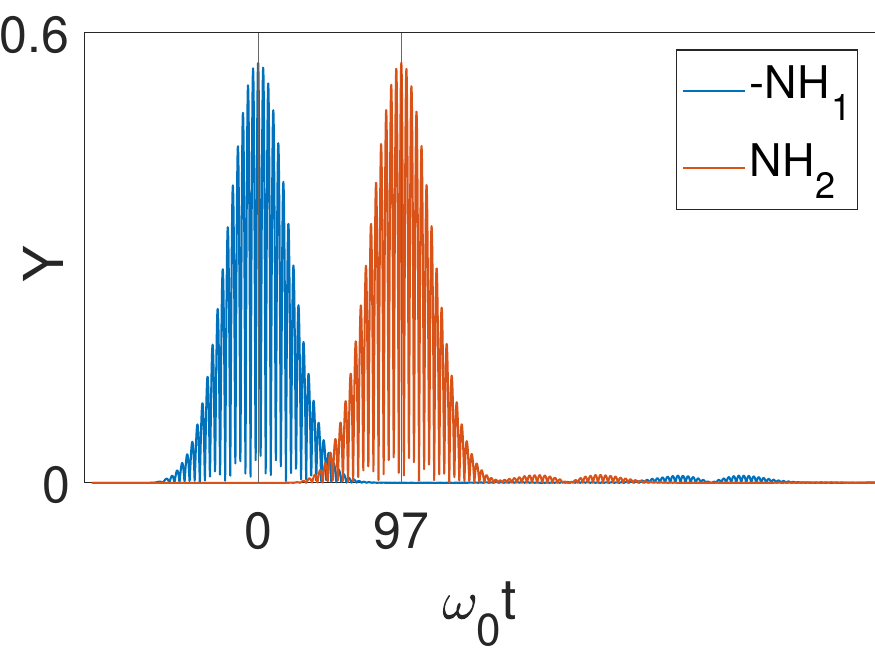} & \includegraphics[height=3.2 cm, valign=c]{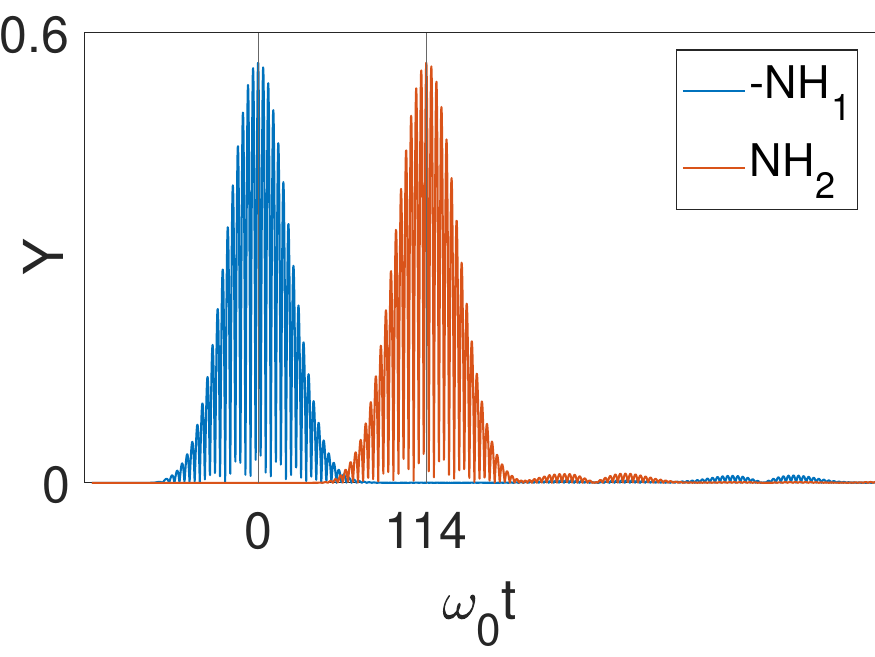} & \includegraphics[height=3.2 cm, valign=c]{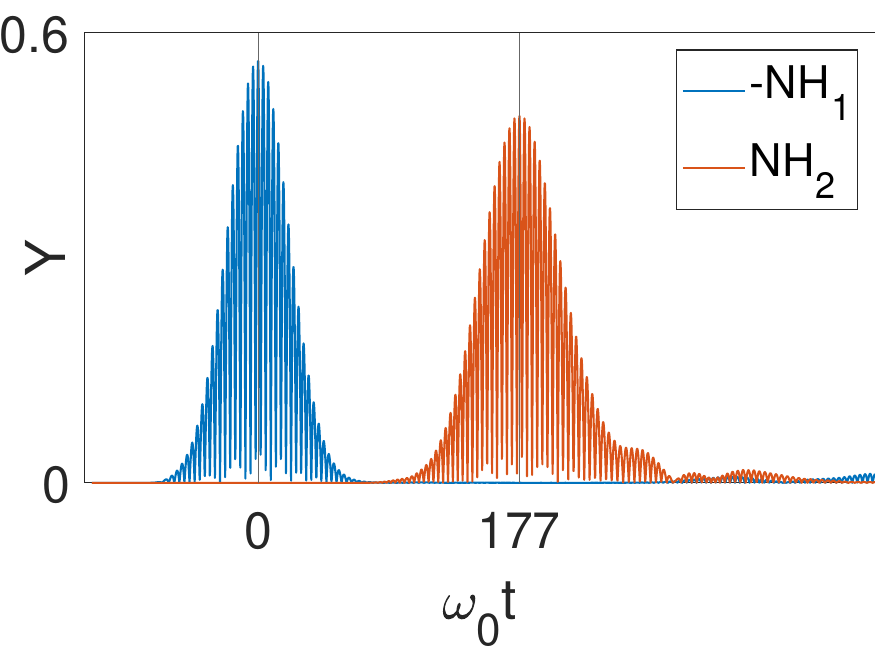}
        \end{tabular}
        \caption{\textbf{The classical model.} (a) Dispersion relation. (b) Reflection coefficient. (c) Decay rate. (d)-(l) Time domain numerical simulations. (d) $\eta$=0. (e)-(h) Unmatched interface. (i)-(l) Matched interface. (e),(i) $\eta$=0.3, $\Omega<\Omega_{g-}$. (f),(j) $\eta$=0.3, $\Omega>\Omega_{g-}$. (g),(k) $\eta$=0.5, $\Omega>\Omega_{g-}$. (h),(l) $\eta$=0.7, $\Omega>\Omega_{g-}$. (m) Group velocity. (n)-(p) Response at the matched interface ends for $\eta$=0.3,0.5,0.7.}
        \label{fig:reflection}
    \end{center}
\end{figure*}
The question that then arises is why the wave that hits the effective barrier at the H1-NH1 boundary is not simply reflected from it. 
To understand this we derive the dispersion relation of an infinite chain governed by \eqref{eq:nonrecip_eq}. Due to the spatial, and not temporal, nature of wave growth/attenuation, we use the convention of complex wavenumber $k=k_R+ik_I$ and real frequency $\omega$. Then, inserting the solution  $y_n(t)\propto q^ne^{i(kna-\omega t)}$ in \eqref{eq:nonrecip_eq} for an infinite blue chain, we obtain
\begin{equation}  \label{eq:dispersion}
  q=\sqrt{\tfrac{1-\eta}{1+\eta}}, \quad  e^{ika}=\tfrac{1}{\beta}\left[1-\tfrac{1}{2}\Omega^2-i\mu\right], \quad \mu=\sqrt{\widehat{\Omega}^2-\eta^2}.
\end{equation}
Here, $\widehat{\Omega}=\Omega\sqrt{1-\Omega^2/4}$, $\Omega=\omega/\omega_0$, $\beta=\sqrt{1-\eta^2}$, and for the red chain the sign of $\eta$ is flipped. The resulting wavenumber $k$ is depicted in Fig. \ref{fig:reflection}(a) for the Hermitian case $\eta=0$, and three non-Hermitian cases $\eta$=0.3,0.5,0.7. 
For $\eta>0$ the real wavenumber has a frequency gap of $\widehat{\Omega}_g=\eta$. This translates to $\Omega_{g-}=\sqrt{2}\sqrt{1-\beta}$, and an upper stopband at $\Omega_{g+}=\sqrt{2}\sqrt{1+\beta}$ (which can be also obtained from solving the open boundary problem of the blue chain \cite{yao2018edge}). 
For $\Omega_{g-}<\Omega<\Omega_{g+}$, we have $k=k_R$, and the waves are propagating with the fixed attenuation/amplification factor $q$. 
The regime at $\Omega<\Omega_{g-}$, for which $k=k_I$, would usually be called evanescent (evanescent to the right but amplified to the left if launched at the middle of the blue chain). 
Therefore, in a standalone non-Hermitian chain the skin effect would hold for all frequencies, but for the tunneling phenomenon the gap plays a crucial role.
This observation can be deduced from $R$, the reflection coefficient from the NH1 section for waves launched in H1. $R$ reads
\begin{equation} \label{eq:reflection_function}
    R=\left[i(\widehat{\Omega}-\mu)-\eta\right]/\left[i(\widehat{\Omega}+\mu)+\eta\right],
\end{equation}
and is depicted in Fig. \ref{fig:reflection}(b).
For $\Omega<\Omega_{g-}$ and $\Omega>\Omega_{g+}$, the function $\mu$, defined in \eqref{eq:dispersion}, is complex-valued, and the magnitude of $R$ is unity for all $\eta>0$.
Therefore, $\Omega=\Omega_{g-}$ is a turning point above which $|R|$ begins to decrease below 1, indicating that tunneling becomes possible. Then $|R|$ sharply increases back to unity toward the upper limit $\Omega=\Omega_{g+}$, and remains unity up to the discrete propagation limit $\Omega=2$ \cite{supplementary}.

We consider two measures to quantify the tunneling. First is the decay rate in the non-Hermitian sections, defined by $q$ in \eqref{eq:dispersion}, indicating how dark is the interface. Second is the transmission rate through the interface, indicating how well the wave is restored after the tunneling. 
The decay rate, depicted in Fig. \ref{fig:reflection}(c), increases with $\eta$, i.e. for a higher $\eta$ the interface is darker. 
The transmission rate, on the other hand, decreases with $\eta$. This can be deduced from the reflection coefficient $R$, which implies that for a higher $\eta$ more is reflected (in a narrower propagation window). 

To illustrate the above, we simulate numerically the time evolution of a Gaussian wavepacket along a generic classical lattice for $\eta$=0.3 at two representative frequencies, $\Omega_1$=0.2 and $\Omega_2$=0.9, for $\eta$=0.5,0.7 at $\Omega_2$, and for $\eta$=0 at $\Omega_1$ for reference. 
Nonreflection was implemented at the leftmost and the rightmost ends of the chain. 
The corresponding time responses are depicted in Fig. \ref{fig:reflection}(d)-(h). 
For $\eta=0$, Fig. \ref{fig:reflection}(d), the chain is fully Hermitian and uniform, and the wavepacket traverses it smoothly, as expected. For $\eta=0.3$, since $\Omega_1$ is below the $\Omega_{g-}$ threshold, and $\Omega_2$ is above, the response, Fig. \ref{fig:reflection}(e)-(f), is fully reflected at $\Omega_1$, but tunnels at $\Omega_2$. 

For $\eta$=0.5 and 0.7, $\Omega$=0.9 is still higher than the respective threshold frequencies 0.52 and 0.76, hence the tunneling occurs, Fig. \ref{fig:reflection}(g)-(h). As $\eta$ grows, less energy penetrates into the interface, rendering it darker. However, due to reflection from the H1-NH1, NH2-H2, and NH1-NH2 junctions, less is transmitted to the other side as well, portraying the decay-transmission trade-off. 
While the exact transmission values can be obtained \cite{supplementary}, we hereby present a method to induce a perfect $|T|=1$ transmission for all frequencies within the propagation window, and for any $\eta$, independently of the decay rate. 
We define control forces $f_{01}$, $f_{12}$, and $f_{20}$ respectively acting at the transition nodes H1-NH1, NH1-NH2, and NH2-H2. Each force obeys the feedback law $f_n=-H_yY_n-H_v\dot{Y}_n$ at the corresponding node $n$, where the control gains read \cite{supplementary}
\begin{equation}
    \begin{array}{c|c|c|c|c}
         & 01,  \Omega<\Omega_{g-}   &  01, \Omega>\Omega_{g-} &  12 & 20  \\
         \hline 
         H^y      & 
             -\eta-i\mu   &  -\eta 
            &  2\eta & -\eta \\
          \hline
         H^v     & \hat{\Omega}/\Omega & (\hat{\Omega}-\mu)/\Omega & 0 & -(\hat{\Omega}-\mu)/\Omega 
    \end{array}.
\end{equation}
The resulting responses are depicted in Fig. \ref{fig:reflection}(i)-(l) for $\eta$=0.3, 0.5, and 0.7, where for $\eta$=0.3 we show both the $\Omega<\Omega_{g-}$ and the $\Omega>\Omega_{g-}$ cases. For $\Omega<\Omega_{g-}$, the wave entirely disappears upon hitting the non-Hermitian interface, whereas for $\Omega>\Omega_{g-}$, total tunneling occurs, independently of $\eta$. 
\begin{figure*}[htpb]
    \centering
 \setlength{\tabcolsep}{3pt}
 \def\arraystretch{1} 
    \begin{tabular}[c]{ccccc}
         \textbf{(a)} & \textbf{(b)} & \textbf{(c)} & \textbf{(d)} &  \\
        \includegraphics[height=2.3 cm, valign=c]{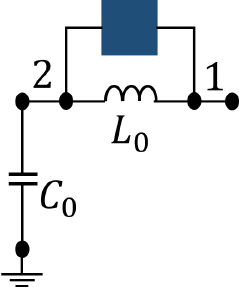} & \includegraphics[height=2.3 cm, valign=c]{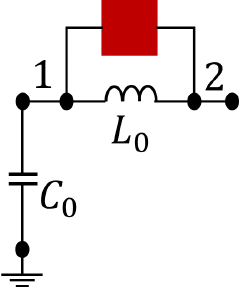} &
        \includegraphics[height=2.7 cm, valign=c]{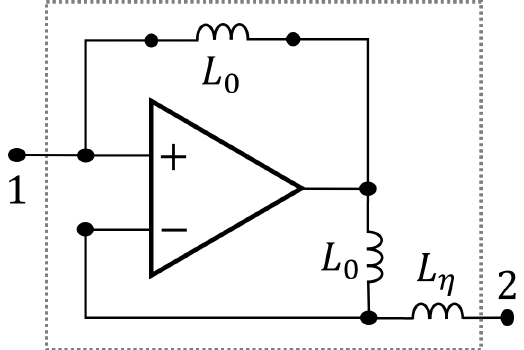} &
   \includegraphics[height=4.4 cm, valign=c]{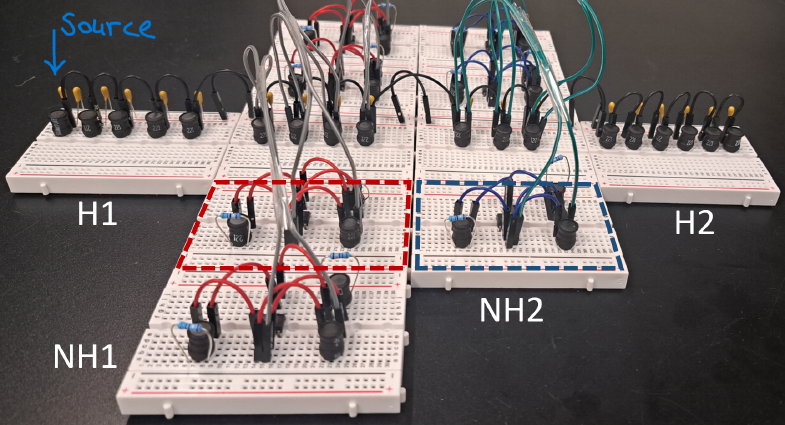}  & \;  \includegraphics[height=3.2cm, valign=c]{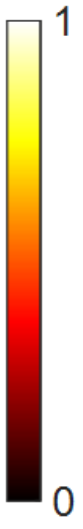}
    \end{tabular} 
\setlength{\tabcolsep}{-1pt}
\def\arraystretch{1.0} 
\begin{tabular}[t]{cccc}
\textbf{(e)} $\bm{\diamond}$, near-full tunneling  & \textbf{(f)} $\bm{\diamond}$, near-full tunneling &  \textbf{(g)} x, near-no tunneling  & \textbf{(h)} x, near-no tunneling \\
    \includegraphics[height=3.4 cm, valign=t]{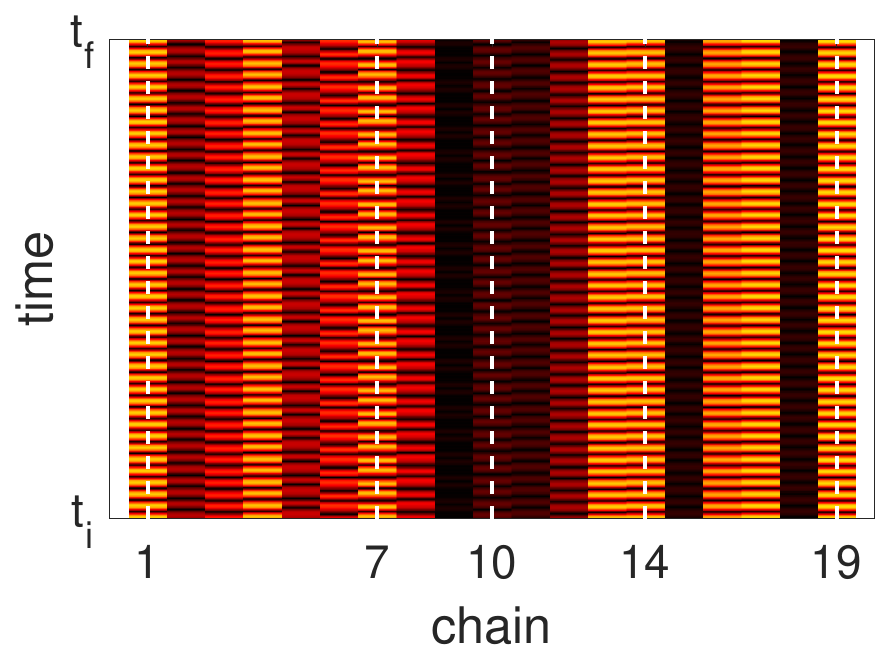} &
    \includegraphics[height=3.4 cm, valign=t]{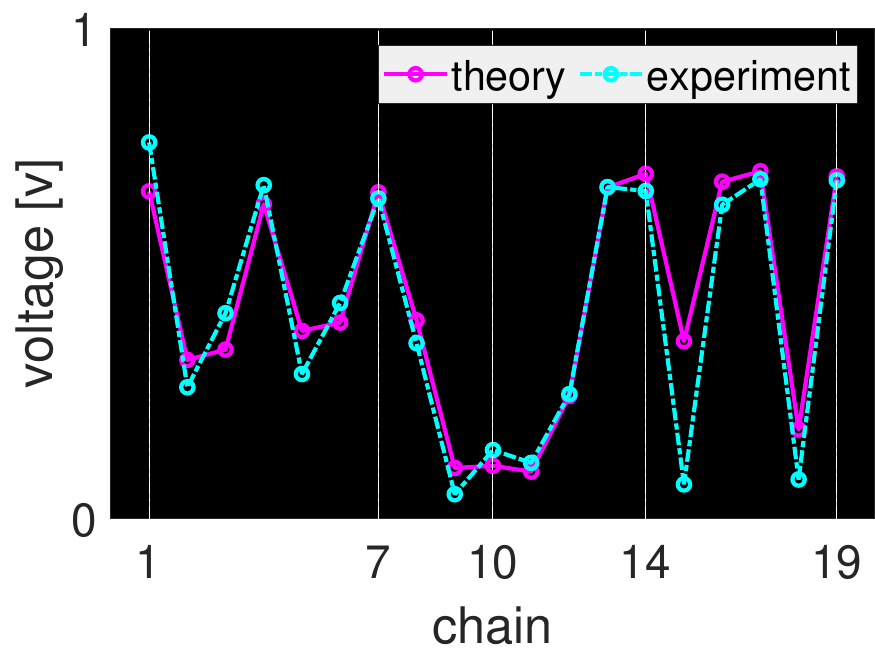} &
    \includegraphics[height=3.4 cm, valign=t]{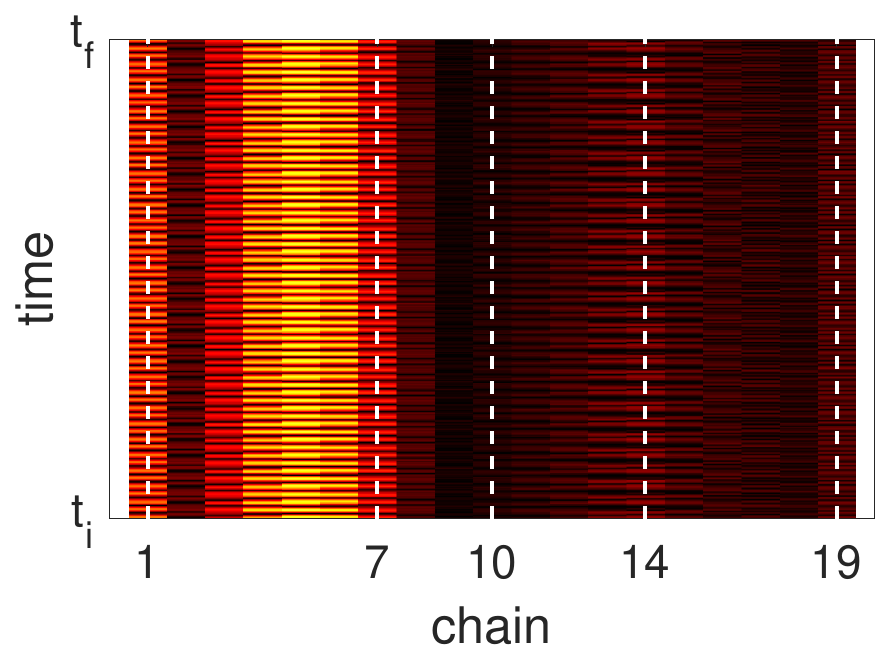} &
    \includegraphics[height=3.4 cm, valign=t]{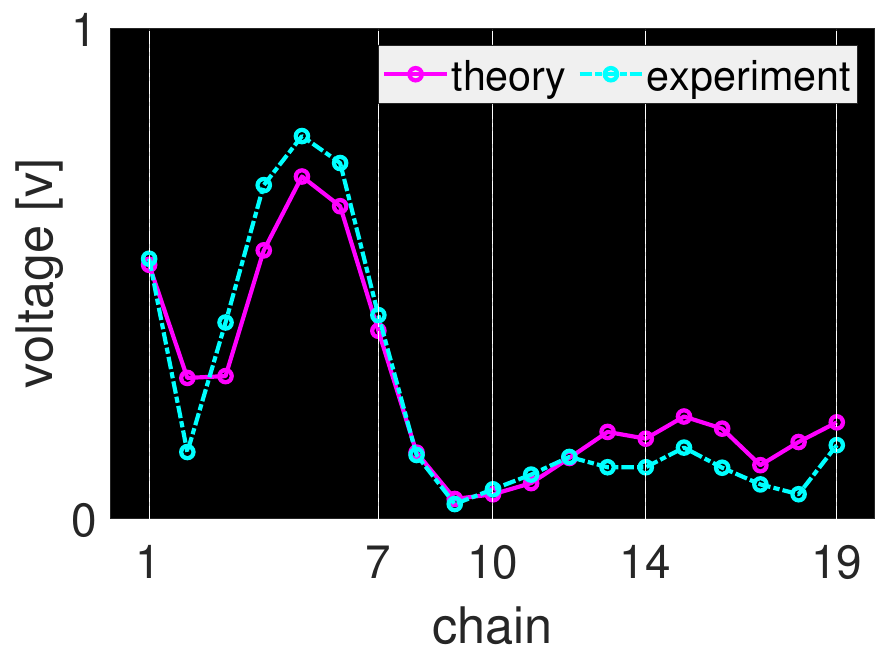}
\end{tabular}
    \caption{\textbf{Experimental demonstration in a topoelectric metamaterial.} (a),(b) The positioning of the active elements, indicating the coupling directionality by the order of terminals 1 and 2. (a) - blue chain, (b) - red chain. (c) Detailed schematic of the active element, generating the required nonreciprocity in the current flow. (d) The experimental setup, featuring a 19 site chain, partitioned according to H1=7, NH1=3, NH2=4, H2=5. The Hermitian sites, wired in black, are assembled along the horizontal breadboards. The $L_0$-$C_0$ components are given by 
    inductors $L_0=220$ muH, and 
    capacitors $C_0=150$ nF. The cells in NH1 and NH2, concatenated vertically, are wired in blue and red, respectively. Each active cell is assembled on a separate breadboard (dashed frames) according to the setup of Fig. \ref{fig:experiment}(a)-(c). It includes an operational amplifier of 10 MHz bandwidth powered by 12 V, and the inductor $L_\eta=440$ muH, leading to $\eta=0.5$. In addition, 
    resistors of $R=30$ Ohm were connected in parallel to the $L_0$ inductors to ensure stability of the operational amplifier. The signal generator impedance was $\approx$100 Ohm, and the source amplitude was 1V. (e),(f) Voltage measurements at the lattice nodes at frequency 28 kHz, (e), and experimental results (cyan) plotted on top of the simulated data (magenta) of the maximal voltage measurements at the 28 kHz, (f), exhibiting the near-full tunneling regime. (g),(h) Same as (e),(f) for 15 kHz, exhibiting the near-no tunneling regime.}
    \label{fig:experiment}
\end{figure*}
The only difference is the propagation speed, which decreases with $\eta$, Fig. \ref{fig:reflection}(m), alongside some dispersion for higher $\eta$. This is exhibited in the spacetime responses, Fig. \ref{fig:reflection}(j)-(l), as a shift in the tunneled beam with respect to the incident. The velocities are derived from the wavepacket arrival times to junctions H1-NH1 and NH2-H2, Fig. \ref{fig:reflection}(n)-(p) for $\eta$=0.3, 0.5, 0.7 ($\eta$=0 not shown), 
and labeled on top of the analytic curves of Fig. \ref{fig:reflection}(m). 

To experimentally demonstrate the phenomenon we consider an electric circuit lattice, also known as topoelectric metamaterial \cite{helbig2020generalized}, in which the sites, the circles in Fig. \ref{fig:model}(a), are given by capacitors $C_0$, and the couplings between the sites, the bars in Fig. \ref{fig:model}(a), are given by inductors $L_0$. 
This capacitor-inductor transmission line constitutes the Hermitian basis of the entire H1-NH1-NH2-H2 chain. 
To induce nonreciprocity in the system, 
we employ embedded active control elements, as illustrated in Fig. \ref{fig:experiment}(a),(b) by squares attached on top of the inductors. These elements actuate the unit cells based on real-time measurements of the responses \cite{geib2021tunable,wen2022unidirectional,wen2023acoustic,nash2015topological,sirota2020active,jana2023gravitational,hofmann2019chiral,zanic2022stability,zhu2023higher}. 
%
The coupling directionality, $1-\eta$ to the left and $1+\eta$ to the right for the blue sites, and the other way around for the red sites, is determined by the order of the terminals connection, 1 and 2.

The active system design is detailed in Fig. \ref{fig:experiment}(c). It's key element is an operational amplifier, which is supplied by constant voltage, and intrinsically stands for both an actuator and a sensor.
It features forward and backward current flow through identical inductors, set here to $L_0$.
The backward flow is fed to the transmission line through $L_\eta$ inductor, rendering negative effective inductance $L_\eta$ between terminals 1 and 2. When positioned in parallel to the main line inductors $L_0$, the amplifier induces nonreciprocity in the transmission line current flow, where $L_\eta=\eta
^{-1}L_0$ determines the nonreciprocity \cite{supplementary}. 
The current flow rate to and from the $n_{th}$ site of a non-Hermitian section becomes 
\begin{equation} \label{eq:nonrecip_current}
    \dot{I}_{n\rightarrow n+1}=(1\pm \eta)\Delta V \; , \; \dot{I}_{n+1\rightarrow n}=(1\mp \eta)\Delta V,
\end{equation}
where the signs order corresponds to the blue and red chain, respectively, and $\Delta V$ is the voltage drop. 
The experiment was carried out using the platform depicted in Fig. \ref{fig:experiment}(d) for $\eta=0.5$. 
We excited the system at the first site of the H1 section and performed two experiments: at frequency 28 kHz, and at 15 kHz, corresponding to $\Omega=1$ and $\Omega=0.54$, respectively portraying a near-full and near-no tunneling regime.
%
The resulting voltage responses for the entire experiment duration
are depicted in Figs. \ref{fig:experiment}(e) and (g). The maximal measurements at each node (cyan) are plotted in Figs. \ref{fig:experiment}(f) and (h) on top of the numerical results (magenta). 
After transmission through the NH1-NH2 interface \cite{endnote1}, in the near-full tunneling regime the voltage in H2 is nearly completely restored to its amplitude of $\approx 0.7$ V in H1, whereas at the near-no tunneling regime the voltage in H2 is restored only to $\approx 0.2$ V. 


To summarize, we demonstrated the non-Hermitian tunneling phenomenon both in quantum and classical systems. 
The quantum systems were assumed noninteracting \cite{endnote2}, as while a many-body extension of the skin effect can be considered\cite{li2023many,gliozzi2024many},
%
the classical analogy exists only in the single-particle case. This is due to the striking similarity between the electronic band-structure of solids and the frequency dispersion of classical waves, despite the difference in the order of the underlying equations\cite{albert2015topological,yang2015topological,sirota2020non}.
This analogy pushes the study of non-Hermitian physics to new regimes, creating new classical waveguiding capabilities.

The non-Hermitian tunneling is essentially different from its Hermitian counterparts due to several reasons. 
First, there is an $\eta$-dependent energy/frequency threshold for the effect to occur.  
Second, the Gaussian wavepacket preserves its width as it passes through the interface, and restores its height at the exit, in contrast to the Hermitian case, where simultaneous width preservation and amplitude reduction would violate the probability conservation. 
Finally, the interface remains dark at all times, unlike Hermitian tunneling, in which waves undergo transmission to higher potential without a void region\cite{katsnelson2006chiral,volovik2016black}.
 
%

Remarkably, the amount of transmitted energy in the non-Hermitian tunneling is independent of the interface length, provided that the skin effect in a standalone nonreciprocal chain is stable. 
In a basic formulation, the transmission rate is traded-off with the interface darkness level, but with dedicated control loops (in the classical case), we managed to obtain a total transmission, independent of $\eta$. 
%
We demonstrated the phenomenon in an active topoelectric metamaterial.
Our platform is modular, as it enables to easily assemble lattices of different sizes, and to create a vast variety of couplings. This includes, e.g., on-site gain and loss \cite{benisty2024controlled}, long range nonreciprocity, nonlinear, and time dependent components, which paves the way to unveil and demonstrate new non-Hermitian phenomena. 


\section*{Acknowledgements}

\textit{This research was supported in part by the Israel Science Foundation Grants No. 2177/23 and 2876/23. 
The authors are grateful to Jensen Li, Badreddine Assouar, and Chen Shen for insightful discussions. 
The authors are especially grateful to Arkadi Rafalovich for his invaluable help with the technical aspects of the experimental setup.}

\bibliographystyle{IEEEtran}

\bibliography{paper}

\begin{thebibliography}{10}
\bibitem{endnote1}
``Since the impedance of the {H}2 right end was not matched (the {H}1 left end was terminated by the signal generator resistance), the response in section {H}2 included reflections from the external end.''
\bibitem{endnote2}
``Although an extensive number of particles in a many-body state cannot be exponentially boundary localized due to the {P}auli exclusion principle, which prevents fermions from clustering at the same site, key features of the single-particle phenomenology persist. {S}pecifically, the eigenstates display a left-right charge density imbalance, and this asymmetry, which tends to move electrons to one edge and holes to the other, has been interpreted as a many-body extension of the skin effect.''
\end{thebibliography}

\onecolumngrid

\appendix

\renewcommand\x{1.6}
\renewcommand\y{3.3}
\renewcommand{\thefigure}{S\arabic{figure}}
\renewcommand{\theequation}{S\arabic{equation}}
\setcounter{equation}{0}


\newpage

\section*{\textbf{Supplementary Material}}

\section{The quantum simulation of Eq. (1), Fig. 1}


Using the quantum model Hamiltonian in Eq. (1), we study the time dynamics of an initial Gaussian quantum wave packet, $\Psi_{in}(x)=(2\pi \delta^2)^{-\frac{1}{4}}\exp(i kx)e^{-(x-x_{0})^{2}/4\delta^{2}}$, of spread $\delta$, which is created in the Hermitian region H1 at position $x_0$ with momentum $k$.
The time evolution of the wavepacket is governed by the Schr\"odinger equation
\begin{equation}
    i\hbar \partial_{t} \Psi(t)=H\Psi(t),
    \label{eq:sq2}
 \end{equation}
where the Hamiltonian $H$ includes all the couplings of the full system including H1-NH1-NH2-H2 in real space. We then solve \eqref{eq:sq2} numerically using state-space formulation with the initial condition $\Psi(x,t=0)=\Psi_{in}(x)$. We choose $k$=1.4$\pi$ and plot the norm of the wavepacket ($\bra{x}\ket{\Psi(t)}$) over position $x$ in Fig. 1(c) and 1(e) of the main text. 
The transmittance $T$ is calculated as  
\begin{equation}    
T=\frac{\sum_{n=1}^{H2}|\Psi_n (t=t_f)|^2}{\sum_{n=1}^{H1}|\Psi_n(t=0)|^2},
\end{equation}
where $\Psi_n$ is the wavefunction at node $n$ of the respective section. For both chains, we observe a consistent value of $T = 0.986$, confirming that the transmittance is independent of the non-Hermitian section length.

\begin{figure}[htpb]
    \centering
    \def\arraystretch{1.1} 
    \begin{tabular}{c c c c}
    \textbf{(a)}  &  \includegraphics[height=\x cm, valign=t]{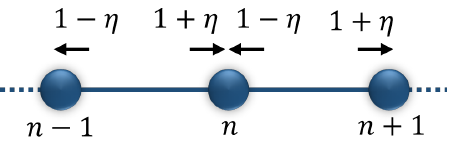}   & \textbf{(c)}  & \includegraphics[height=\x cm, valign=t]{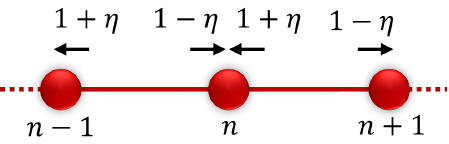} \\
  \textbf{(b)}  &  \includegraphics[height=4.0 cm, valign=t]{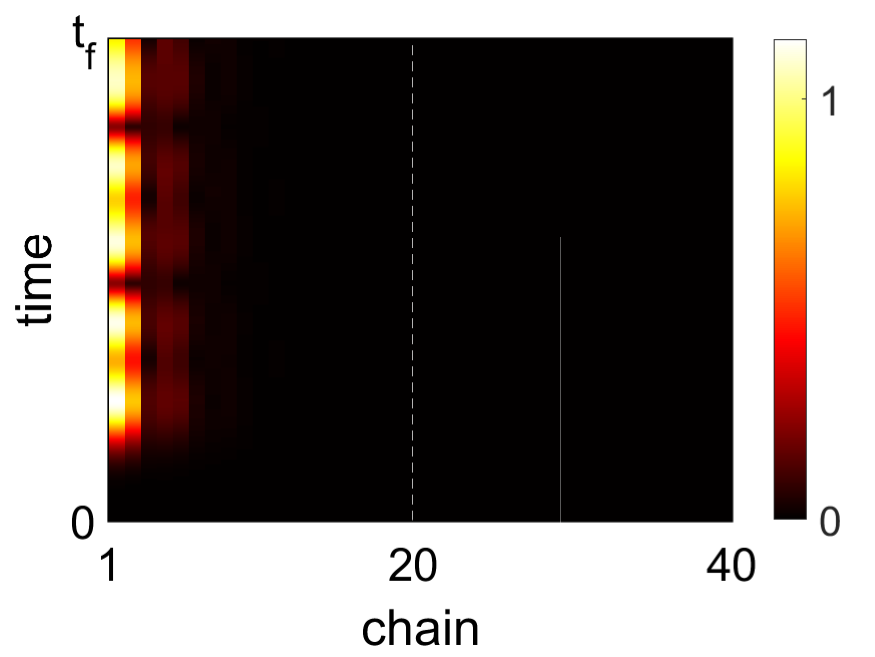}     &  \textbf{(d)}  & \includegraphics[height=4.0 cm, valign=t]{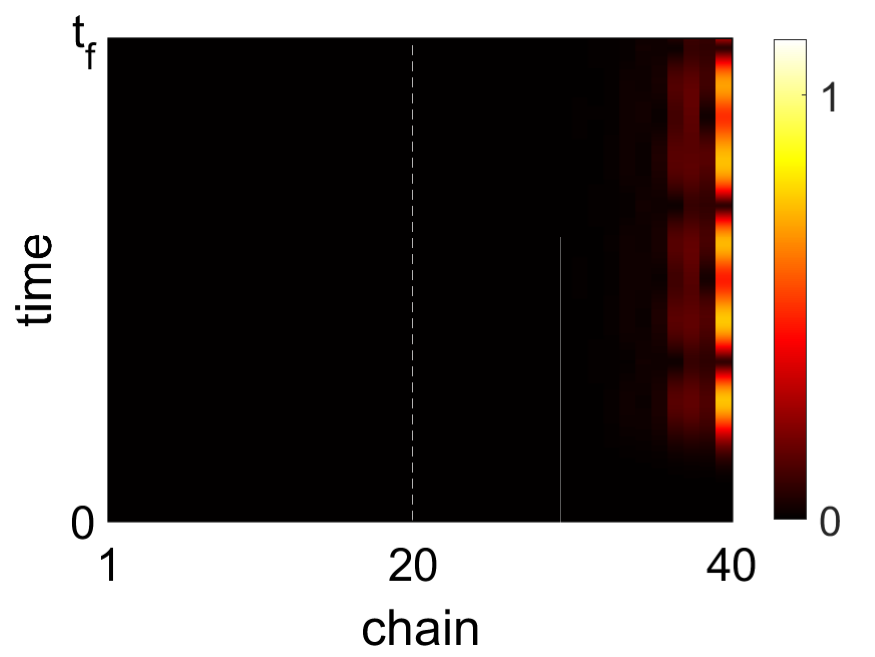}
    \end{tabular}
    \caption{The non-Hermitian skin effect as a function of nonreciprocity direction in the quantum system. (a),(b) [(c),(d)] The creation operator is coupled to the annihilation operator to the right with a stronger coupling, and thus the skin effect is stronger in the left [right] direction.}
    \label{fig:left_right_skin}
\end{figure}

To illustrate the convention of stronger and weaker coupling due to the nonreciprocity, we perform numerical simulations of the standalone non-Hermitian sections NH1 and NH2, and a source term at the chain center. In NH1, the blue chain in Fig. \ref{fig:left_right_skin}(a), the electron creation operator at each site is coupled to the annihilation operator of its nearest neighbor to the right with a stronger coupling of $1+\eta$ and to the left with a weaker coupling of $1-\eta$, indicating that the electron has a stronger hopping strength towards the left. This results in the accumulation of skin modes at the left edge, Fig. \ref{fig:left_right_skin}(b). The situation is flipped for the red NH2 chain in Fig. \ref{fig:left_right_skin}(c), where the skin modes accumulate at the right boundary, as depicted in Fig. \ref{fig:left_right_skin}(d). An analytical derivation of a similar system appears, for example, in Ref. [32] of the main text.

\begin{figure}[htpb]
\begin{center}
\setlength{\tabcolsep}{10pt}
\def\arraystretch{1.1} 
\begin{tabular}[c]{ccc}
\textbf{(a)} & \textbf{(b)} & \textbf{(c)}  \\
    \includegraphics[height=\x cm, valign=c]{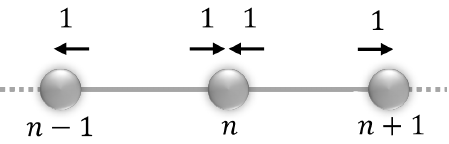} &
    \includegraphics[height=\x cm, valign=c]{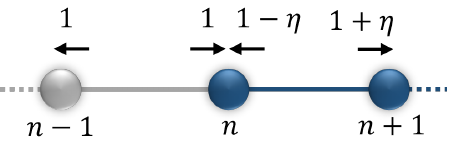} &
    \includegraphics[height=\x cm, valign=c]{Figures_sup/classical_model_NH1_NH1.pdf} \\
    \textbf{(d)} & \textbf{(e)} & \textbf{(f)}  \\
    \includegraphics[height=\x cm, valign=c]{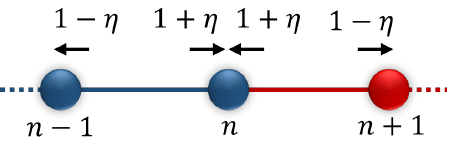} &
    \includegraphics[height=\x cm, valign=c]{Figures_sup/classical_model_NH2_NH2.pdf} &
    \includegraphics[height=\x cm, valign=c]{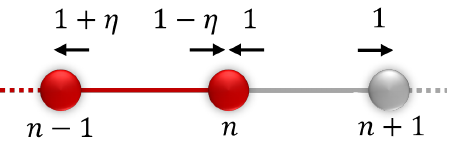} 
\end{tabular}
\end{center}
\caption{The various segments in the H1-NH1-NH2-H2 chain of Fig. 1(a) of the main text.}
\label{fig:segments}
\end{figure}

\section{The equivalent classical model $\Ddot{\textbf{Y}}=D\textbf{Y}$, Eq. (2)}

We consider a generic mass-spring chain with masses $M_0$ connected by springs of nominal stiffness $K_0$, and ranging between $K_0(1\pm\eta)$, where $\eta\in(0,1)$.
For the various segments of the chain in Fig. 1(a) of the main text, as detailed in Fig. \ref{fig:segments}, we obtain the following equations of motion:
\begin{subequations} \label{eq:segments_eq}
\begin{align}
        M_0\Ddot{Y}_n&=K_0((Y_{n+1}-Y_n)-(Y_n-Y_{n-1}))=K_0(Y_{n+1}-2Y_n+Y_{n-1})\\
        M_0\Ddot{Y}_n&=K_0((1+\eta)(Y_{n+1}-Y_n)-(Y_n-Y_{n-1}))=K_0((1+\eta)Y_{n+1}-(2+\eta)Y_n+Y_{n-1}) \label{eq:S3b} \\
        M_0\Ddot{Y}_n&=K_0((1+\eta)(Y_{n+1}-Y_n)-(1-\eta)(Y_n-Y_{n-1}))=K_0((1+\eta)Y_{n+1}-2Y_n+(1-\eta)Y_{n-1}) \label{eq:nonrecip_mechanical_blue}\\
        M_0\Ddot{Y}_n&=K_0((1-\eta)(Y_{n+1}-Y_n)-(1-\eta)(Y_n-Y_{n-1}))=K_0(1-\eta)(Y_{n+1}-2Y_n+Y_{n-1}) \label{eq:S3d} \\
        M_0\Ddot{Y}_n&=K_0((1-\eta)(Y_{n+1}-Y_n)-(1+\eta)(Y_n-Y_{n-1}))=K_0((1-\eta)Y_{n+1}-2Y_n+(1+\eta)Y_{n-1}) \label{eq:nonrecip_mechanical_red}\\
        M_0\Ddot{Y}_n&=K_0((Y_{n+1}-Y_n)-(1+\eta)(Y_n-Y_{n-1}))=K_0(Y_{n+1}-(2+\eta)Y_n+(1+\eta)Y_{n-1})  \label{eq:S3f}
\end{align}
\end{subequations}
Assembling the equations in \eqref{eq:segments_eq} into a matrix for a finite chain gives the dynamical matrix $D$. Equations (S1c) and (S1e) equal Eq. (2) of the main text (with a flipped sign of $\eta$ for (S1e)).

\section{The reflection coefficient, dispersion relation, and group velocity for the quantum and classical models}

\begin{figure}[htpb]
\begin{center}
\setlength{\tabcolsep}{1pt}
\begin{tabular}{cccc}
    \textbf{(a)} & \textbf{(b)} & \textbf{(c)} & \textbf{(d)} \\
    \includegraphics[height=2.1 cm, valign=c]{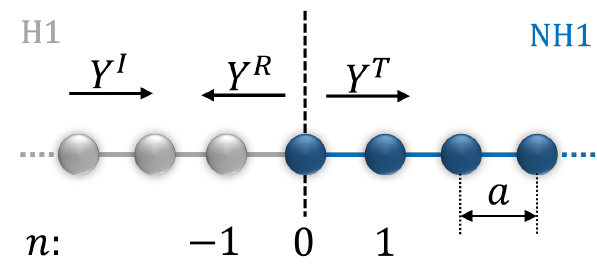} & \includegraphics[height=3.1 cm, valign=c]{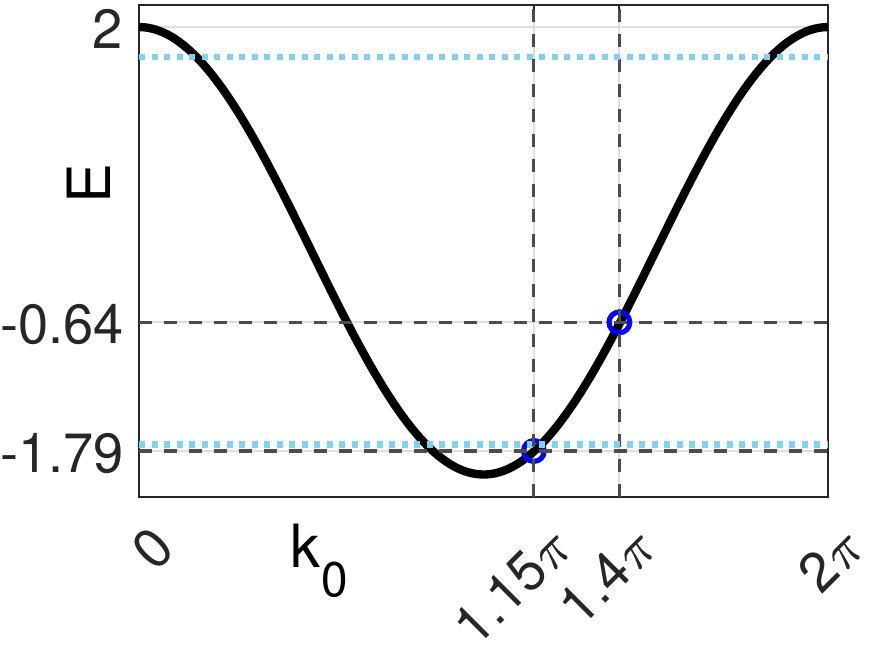}
&  
\includegraphics[height=3.2 cm, valign=c]{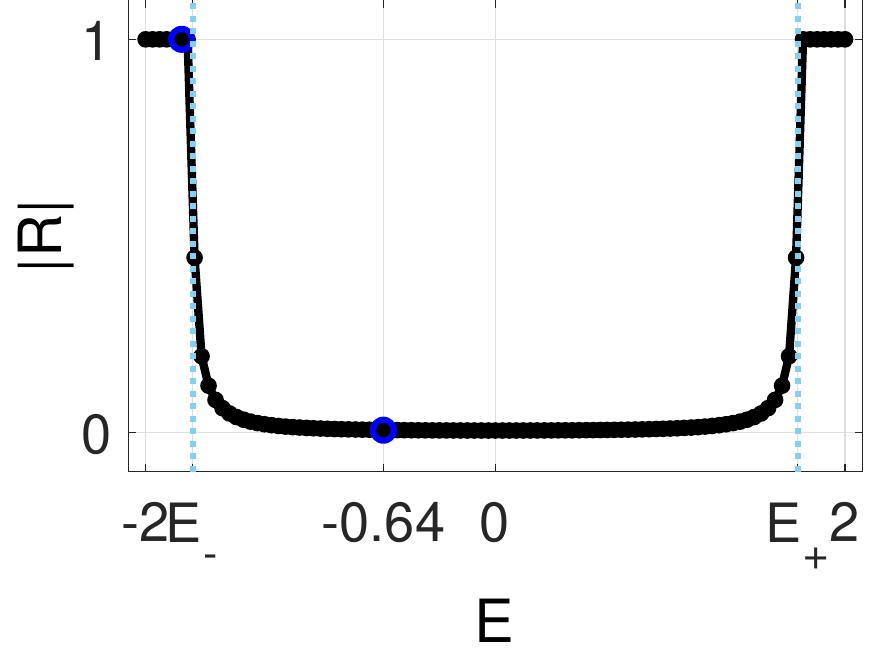} 
&  
\includegraphics[height=3.2 cm, valign=c]{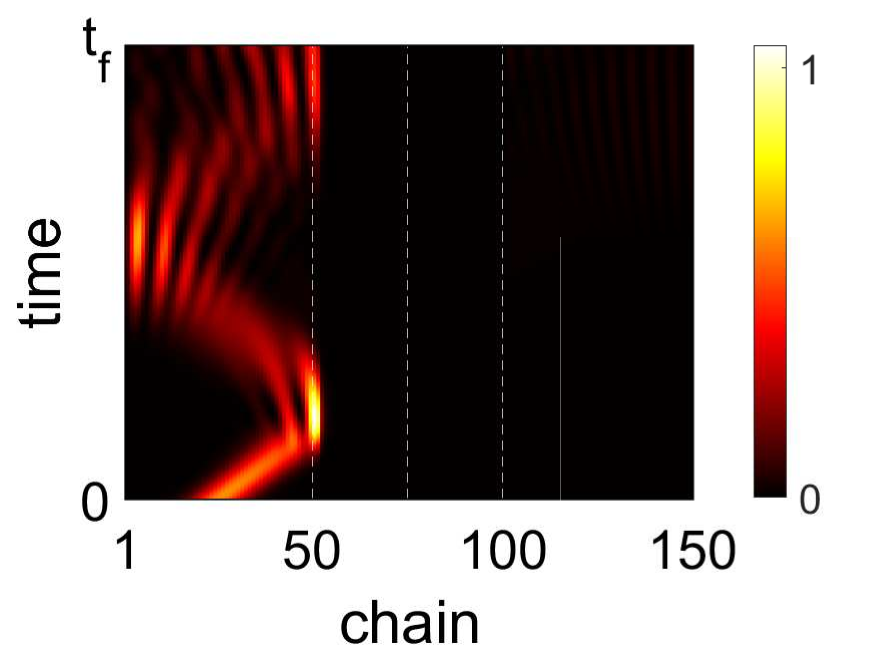} 
\end{tabular}
\end{center}
\caption{(a) The reflection schematic (classical). (b)-(d) Quantum: (b) The dispersion relation. (c) The reflection coefficient for $\eta=0.5$. (d) Time-domain simulation of a wave packet with an initial momentum $k_{0}=1.15\pi$, and associated energy E=-1.794 (below the threshold).}
\label{fig:reflection_discrete}
\end{figure}

\textbf{The classical model.} The reflection schematic is depicted in Fig. \ref{fig:reflection_discrete}(a) (equivalent to Fig. \ref{fig:segments}(b)). We define the solution in section H1 as the sum of an incidence wave $Y^I_n(t)=e^{i\omega t}\overline{Y}^I_n$ and a reflected wave $Y^R_n(t)=e^{i\omega t}\overline{Y}^R_n$, and in section NH1 as a transmitted wave $Y^T_n(t)=e^{i\omega t}\overline{Y}^T_n$. 
Excluding the time harmonic factor we thus obtain that for $n<0$ the response is given by $\overline{Y}_n=\overline{Y}^I_n+\overline{Y}^R_n$, whereas for $n>0$ by $\overline{Y}_n=\overline{Y}^T_n$.
Assuming both sections of infinite length for the calculation, we employ Bloch wave solution
\begin{equation}  \label{eq:Y_IRT}
    \overline{Y}^I_n=e^{ik_0na} \quad , \quad \overline{Y}^R_n=Re^{-ik_0na} \quad , \quad \overline{Y}^T_n=Tq^ne^{ikna},
\end{equation}
where $R$ and $T$ are respectively the reflection and transmission functions between the Hermitian and non-Hermitian medium, and $q=\sqrt{\frac{1-\eta}{1+\eta}}$ is the decay function, defined in Eq. (3) of the main text. $k_0$ and $k$ are the wavenumbers in the Hermitian and non-Hermitian mediums, respectively, and $a$ is the lattice constant.
Therefore, at the interface $n=0$ the solution reads
\begin{equation} \label{eq:cont_1} \overline{Y}_0=\overline{Y}^I_0+\overline{Y}^R_0=\overline{Y}^T_0 \quad \Rightarrow \quad 1+R=T.
\end{equation}
This is the expected continuity condition, similar to the one obtained for Fresnel coefficients in continuous media.
Employing the explicit equation of motion at $n=0$, 
\begin{equation}
    M_0\Ddot{Y}_0(t)=K_0((1+\eta)(Y_1(t)-Y_0(t))-(Y_0(t)-Y_{-1}(t)))    
\end{equation}
we obtain the second continuity condition
\begin{equation}
    (2+\eta-\Omega^2)\overline{Y}_0=(1+\eta)\overline{Y}_1+\overline{Y}_{-1},
\end{equation}
where $\Omega=\omega/\omega_0$ and $\omega_0^2=K_0/M_0$.
Substituting the solution $\overline{Y}_0=1+R$, $\overline{Y}_1=Tqe^{ika}$, and $\overline{Y}_{-1}=\overline{Y}_{-1}^I+\overline{Y}_{-1}^R=e^{-ik_0a}+Re^{ik_0a}$, together with the first continuity condition, we solve for $R$, as
\begin{equation}  \label{eq:R_eka}
    R=\frac{-(2+\eta-\Omega^2)+e^{-ik_0a}+(1+\eta)qe^{ika}}{(2+\eta-\Omega^2)-e^{-ik_0a}-(1+\eta)qe^{ika}}. 
\end{equation}
To obtain the expressions for $k_0$ and $k$, we calculate the dispersion relation in each medium. For $n>0$ (Eq. \eqref{eq:nonrecip_mechanical_blue}) it is given by
\begin{equation}  \label{eq:S9}
    (1+\eta)qe^{ika}+(\Omega^2-2)+(1-\eta)q^{-1}e^{-ika}=\beta e^{ika}+(\Omega^2-2)+\beta e^{-ika}=0,
\end{equation}
where $\beta=\sqrt{1-\eta^2}$.
Solving a quadratic equation for $e^{ika}$ gives
\begin{equation}  \label{eq:eka}
    e^{\pm ika}=\left(1-\tfrac{1}{2}\Omega^2\mp i\mu\right)/\beta, \quad \mu=\sqrt{\widehat{\Omega}^2-\eta^2}, \quad e^{\pm ik_0a}=1-\tfrac{1}{2}\Omega^2\mp i\widehat{\Omega},
\end{equation}
and $\widehat{\Omega}=\Omega\sqrt{1-\Omega^2/4}$, as was shown in the main text. 
Substituting \eqref{eq:eka} into \eqref{eq:R_eka}, we obtain the reflection function presented in Eq. (4) of the main text. 
Solving the inequality $\eta^2-\widehat{\Omega}^2>0$, or, explicitly $\Omega^4-4\Omega^2+4\eta^2<0$, implies the statement of the main text that for $\Omega<\Omega_{g-}$ and $\Omega>\Omega_{g+}$, where $\Omega_{g\mp}=\sqrt{2}\sqrt{1\mp\beta}$, $|R|$ is unity for all $\eta$. 

To obtain the group velocity of Fig. 2(m), we extract the real part of \eqref{eq:eka}, as $\cos{ka}=(1-\Omega^2/2)/\beta$. Rearranging and calculating $\partial\Omega/\partial ka$, reads
\begin{equation}
    v_g=\frac{1}{\sqrt{2}}\frac{\beta\sin{ka}}{\sqrt{1-\beta\cos{ka}}}.
\end{equation}

\textbf{The quantum model.} Replacing  in \eqref{eq:Y_IRT} the frequency $\omega$ by the energy $E$, and the classical variable $Y$ by the quantum wavefunction $\Psi$, we obtain the same first continuity condition as in \eqref{eq:cont_1}. 
For the second condition we employ the real-space Schr\"odinger equation at $n=0$, which reads
\begin{equation}
    i \partial_{t} \Psi_{0}^{}(t)=(1+\eta)\Psi_{1}+\Psi_{-1},
    \label{eq:sq5}
 \end{equation}
and together with \eqref{eq:Y_IRT}-\eqref{eq:cont_1} leads to the reflection coefficient
\begin{equation}  \label{eq:R_12}
    R=\frac{e^{-i k_{0}a} + qe^{i k a} (1+\eta) -E}{-e^{i k_{0}a} - qe^{i k a} (1+\eta)+E},
\end{equation}
where the decay function $q$ is the same as in the classical case. Now, for $n > 0$, we obtain 
  \begin{equation}
    E = (1+\eta)qe^{ika}+(1-\eta)q^{-1}e^{-i k a},
    \label{eq:sq7}
 \end{equation}
which is the same as \eqref{eq:S9}, just without the classical restoring forces.
The solutions read $e^{\pm i k a}=\frac{E\mp \sqrt{4(\eta^2 -1)+E^2}}{2\beta}$ and $e^{\pm i k_{0} a}=\frac{E\mp \sqrt{E^2-4}}{2}$. 
Similarly, the expression for the reflection coefficient $R_{mid}$ between the NH1 and NH2 sections can be obtained as
 \begin{equation}  \label{eq:R_mid}
  R_{mid}=\frac{-q(e^{-i k a}+e^{i k_{2} a})(1-\eta)+ E q^2}{ (qe^{i k a}+q^{-1}e^{-i k_{2} a}) (1-\eta)-E}
 \end{equation}
 where $e^{\pm i k_{2} a}=\frac{E\mp \sqrt{4(\eta^2 -1)+E^2}}{2\beta}$ and $k_{2}$ denotes the wave number associated with section NH2. By substituting the coefficients $\bigl\{  e^{\pm i k a},e^{\pm i k_{2} a} \bigl\}$ into \eqref{eq:R_mid}, we find that
 \begin{equation}
     R_{mid}=0,
 \end{equation}
indicating that the wavepacket is transmitted between the NH1 and NH2 sections without reflection. This is distinct from the classical system due to the restoring forces in the latter.
$|R|$ of \eqref{eq:R_12} is depicted in Fig. \ref{fig:reflection_discrete}(c) for $\eta=0.5$ as a function of $E$. Two critical values are identified: $E_{\pm}$=$\pm 2\beta$, above which $|R|$=1. In Fig. 1(c) of the main text, we observe that the wave packet associated with E=-0.64, which is above the threshold energy, tunnels through the interface with negligible reflection, yielding $R \approx $0.006, consistent with the analytical value. In contrast, a wave packet corresponding to E=-1.794, which exceeds the critical threshold of $\pm 2\beta$, undergoes total reflection from the H1-NH1 interface, as confirmed by the numerical simulation in Fig. \ref{fig:reflection_discrete}(d). 

\section{Unmatched tunneling}

\begin{figure*}[htpb]
\begin{center}
\begin{tabular}{cc}
\setlength{\tabcolsep}{0pt}
\def\arraystretch{0.9} 
\begin{tabular}[t]{cc}
\textbf{(a)} $\bm{\square}$ & \textbf{(b)} $\bm{*}$ \\ 
    \includegraphics[height=3.2 cm, valign=c]{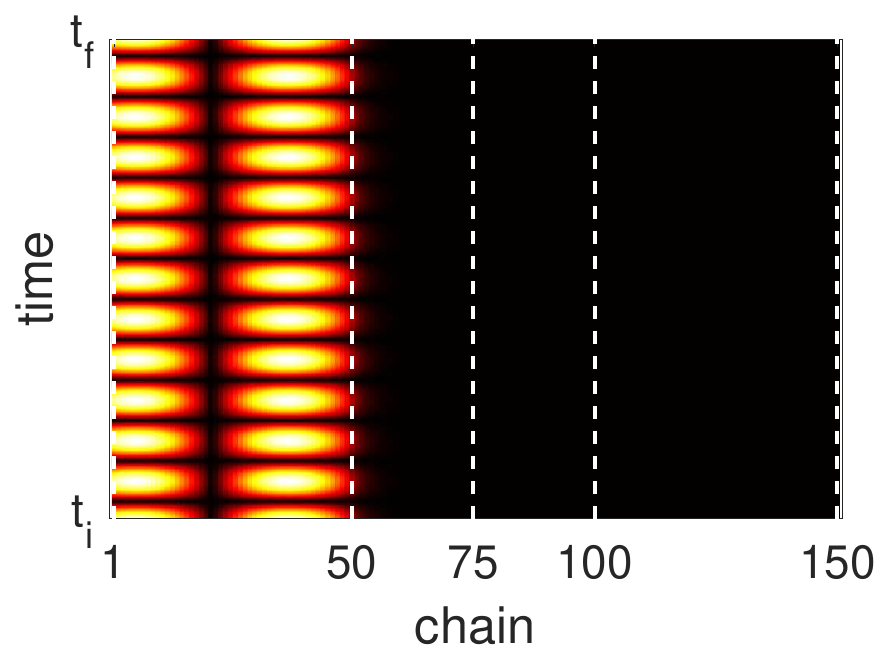} &  \includegraphics[height=3.2 cm, valign=c]{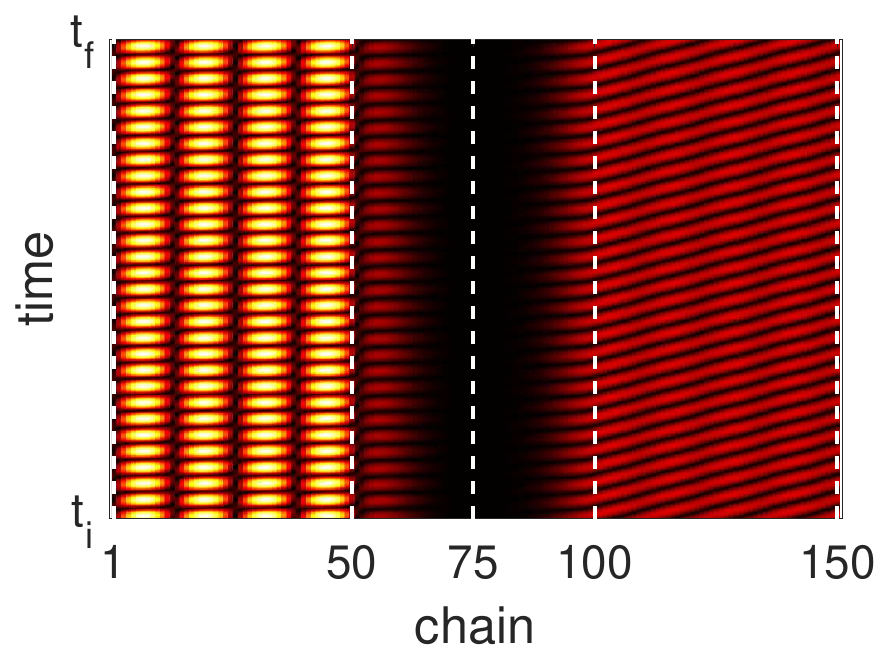}\\ \textbf{(c)} $\bm{\circ}$ & \textbf{(d)} $\bm{\diamond}$  \\     
    \includegraphics[height=3.2 cm, valign=c]{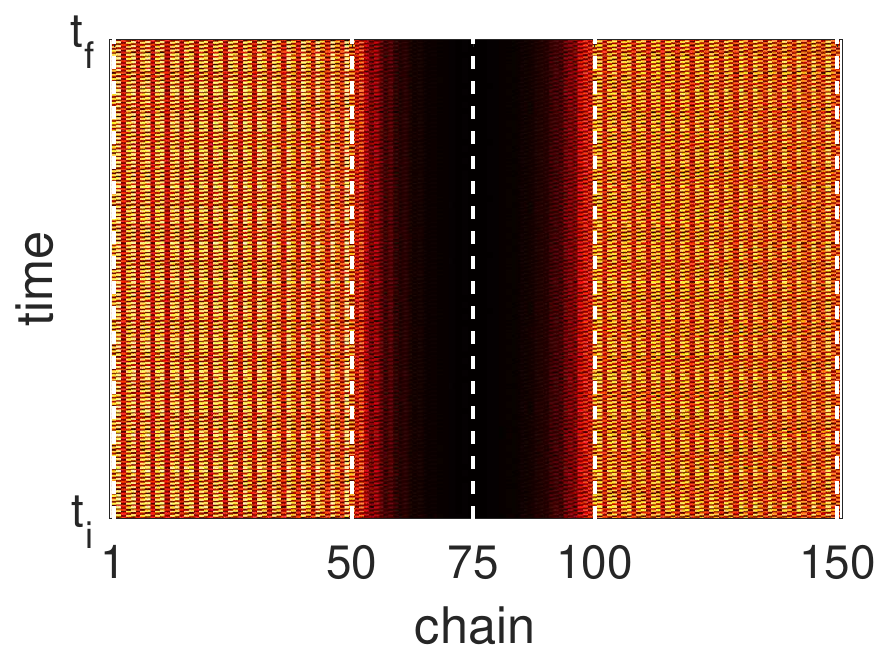} &
    \includegraphics[height=3.2 cm, valign=c]{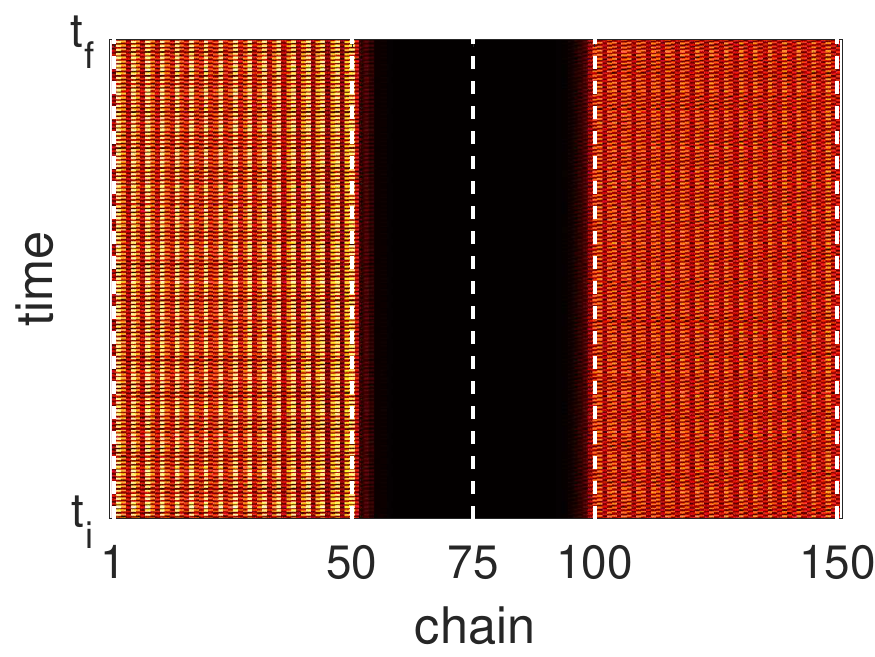}  
\end{tabular} &  \def\arraystretch{1.8}  \begin{tabular}[t]{c}
\textbf{(e)} \\
     \includegraphics[height=4.8 cm, valign=c]{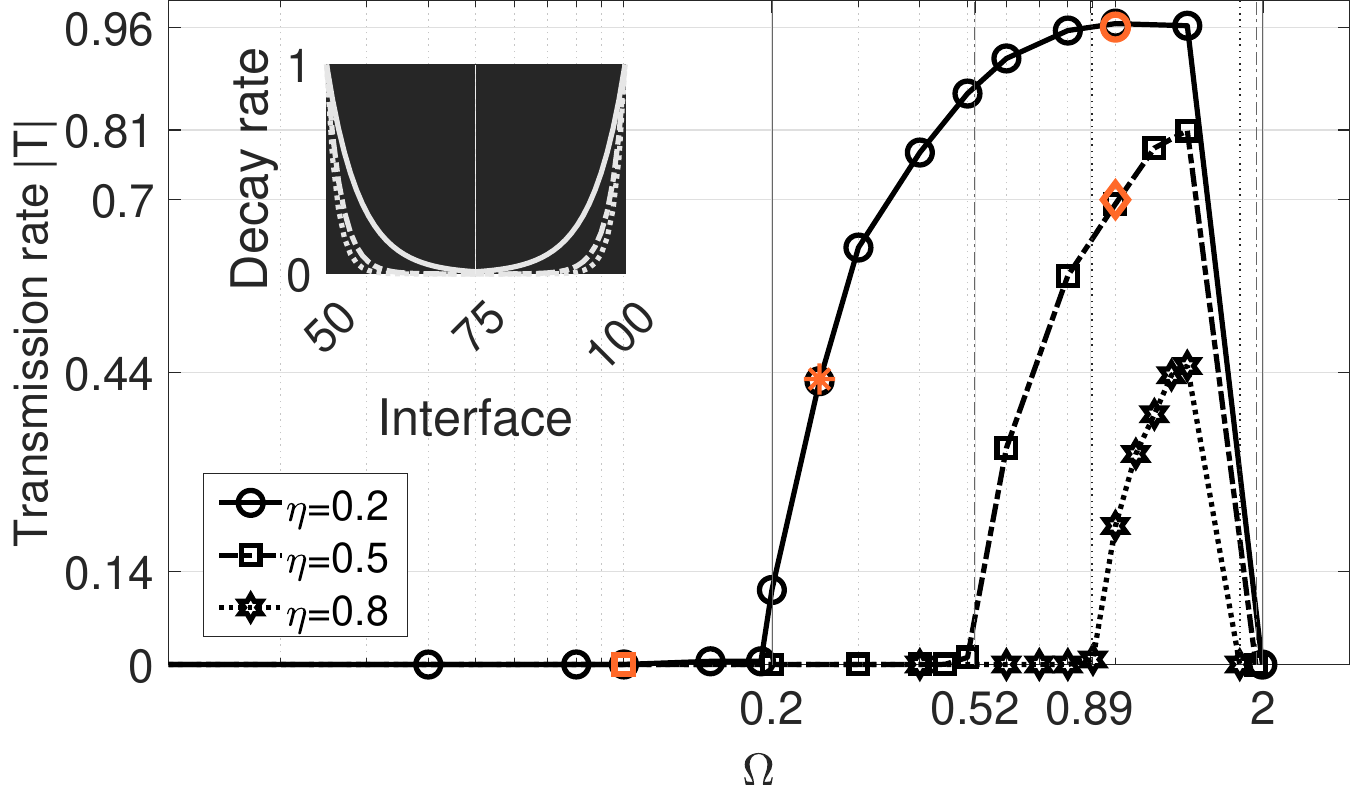}  \\
     \includegraphics[height=0.3 cm]{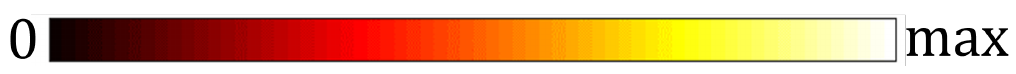}
\end{tabular}   
\end{tabular}
\end{center}
\caption{\textbf{Unmatched tunneling: transmission and decay rates trade-off demonstration.} (a)-(d) Time domain numerical simulations, corresponding to the orange square, star, circle and diamond working points labeled in panel (e). Impedance matching was implemented at sites 1 and 150.
(e) The transmission function $|T|$, calculated numerically for $\eta=0.2$ (solid-circles), $\eta=0.5$ (dashed-squares) and $\eta=0.8$ (dotted-hexagrams), and the corresponding decay rate in the inset. 
$|T|$ is zero below $\Omega_{g-}$ and above $\Omega_{g+}$.}
\label{fig:reflection_trade_off}
\end{figure*}

To illustrate the transmission and decay rates trade-off in the unmatched tunneling scenario, we consider three representative frequencies, $\Omega=0.1$, $0.25$ and $1$. For $\eta=0.2$, $|T|$ at these frequencies is labeled in Fig. \ref{fig:reflection_trade_off}(e) by an orange square, star and circle. 
For each of these working points we simulate numerically the time evolution of a harmonic wave along a generic classical lattice of 150 sites, sectioned as 50-25-25-50. The corresponding time responses are depicted in Fig. \ref{fig:reflection_trade_off}(a)-(d), respectively yielding 0, 44 and 96 percent of tunneling. 
For comparison, we simulate the $\Omega=1$ case also for $\eta=0.5$, labeled by a diamond in Fig. \ref{fig:reflection}(e).
The resulting time response is given in Fig. \ref{fig:reflection}(d) with 70 percent of tunneling, but a considerably higher decay rate than in the $\eta=0.2$ case.
It can be observed that for a given $\eta$, the higher is the frequency, the higher is the transmission magnitude, i.e. the more energy is tunneled through the interface. 
The transmission, however, is limited by $\eta$-dependent constant at high frequencies, given by 96\%, 81\% and 44\% for the selected $\eta$. The transmission rate increases as $\eta$ decreases, approaching 1 for a zero $\eta$. 
The decay rate, on the contrary, increases together with $\eta$ (therefore, unity transmission simply means that the chain is uniform and no tunneling takes place).
This implies that the tunneling quality is a trade-off between the transmission and the decay rates.
We then obtain $|T|$ numerically by propagating wavepackets from H1 through NH1-NH2, and calculating its $L^2$ norm in the H2 section, as depicted in Fig. \ref{fig:reflection_trade_off}(e) for $\eta=0.2$, $\eta=0.5$ and $\eta=0.8$. 
The calculation is terminated at the tunneling upper bound $\Omega=\Omega_{g+}$.
For $\Omega\leq\Omega_{g-}$ and $\Omega\geq\Omega_{g+}$, $|T|=0$ holds. 
The corresponding decay rate, which is frequency independent, is added in the inset.

\section{Impedance matching of the interface, Fig. 2 (i)-(l), Eq. (5).}

We begin with the transition H1$\rightarrow$NH1, defining this node by $n=0$.
Accounting for the control forces, the governing equation at $n=0$, \eqref{eq:S3b}, becomes
\begin{equation}  \label{eq:R_01_eq}
    M_0\Ddot{Y}_0=K_0((1+\eta)(Y_1-Y_0)-(Y_0-Y_{-1}))+f_{01}, \quad f_{01}=-K_0H_yY_0-\sqrt{M_0K_0}H_v\dot{Y}_0.
\end{equation}
Repeating the procedure of \eqref{eq:Y_IRT}-\eqref{eq:eka}, we obtain 
\begin{equation}
    -\Omega^2\overline{Y}_0=(1+\eta)(\overline{Y}_1-\overline{Y}_0)-(\overline{Y}_0-\overline{Y}_1)-H_y\overline{Y}_0-i\Omega H_v\overline{Y}_0,
\end{equation}
and the reflection coefficient takes the form
\begin{equation}  \label{eq:R_01}
 \hat{\Omega}>\eta: \quad   R_{01}=\frac{i(\hat{\Omega}-\sqrt{\hat{\Omega}^2-\eta^2}-\Omega H_v)-\eta-H_y}{i(\hat{\Omega}+\sqrt{\hat{\Omega}^2-\eta^2}+\Omega H_v)+\eta+H_y}, \quad \hat{\Omega}<\eta: \quad   R_{01}=\frac{i(\hat{\Omega}-\Omega H_v)-\eta-H_y-\sqrt{\eta^2-\hat{\Omega}^2}}{i(\hat{\Omega}+\Omega H_v)+\eta+H_y+\sqrt{\eta^2-\hat{\Omega}^2}}.
\end{equation}
The values of the control gains $H_y$ and $H_v$ in Eq. (5) of the main text take $R_{01}$ to zero.
Similarly, for the transition NH1$\rightarrow$NH2 we invoke \eqref{eq:S3d} with the control force $f_{12}$, and obtain
\begin{equation}
    M_0\Ddot{Y}_0=K_0(1-\eta)(Y_1-2Y_0+Y_{-1})+f_{12}, \quad f_{12}=-K_0H_yY_0-\sqrt{M_0K_0}H_v\dot{Y}_0.
\end{equation}
This leads to
\begin{equation}  \label{eq:R_12_c}
    R_{12}=R_{mid}=\frac{\Omega^2-2(1-\eta)-H_y+H_vi\Omega+(1-\eta)\left(e^{ika}+e^{-ika}\right)}{\Omega^2-2(1-\eta)+H_y+H_vi\Omega+(1-\eta)2e^{ika}}, 
\end{equation}
where $e^{\pm ika}$ is defined in \eqref{eq:eka}, and to the corresponding gains in Eq. (5) of the main text. The zero velocity gain in this case indicates that the impedance mismatch between these two chains with mirrored but otherwise identical couplings was solely due to the restoring forces of the springs (which are absent in the quantum case, rendering the quantum $R_{mid}$ in \eqref{eq:R_mid} as zero). 
For the NH2$\rightarrow$H2 transition we cannot use $R_{01}$ in \eqref{eq:R_01} because of nonreciprocity, i.e. the transmission from H1 to NH1 is not exactly the same as from NH1 to H1, or, equivalently, from NH2 to H2. Invoking \eqref{eq:S3f} and following similar steps to \eqref{eq:R_01_eq}-\eqref{eq:R_01}, gives 
\begin{equation}
    R_{10}=\frac{i(-\hat{\Omega}+\sqrt{\hat{\Omega}^2-\eta^2}-\Omega H_v)-\eta-H_y}{i(\hat{\Omega}+\sqrt{\hat{\Omega}^2-\eta^2}+\Omega H_v)+\eta+H_y},
\end{equation}
and the respective gains in Eq. (5) of the main text. We note that $R_{10}$ does not equal $R_{10}$ with a flipped $\eta$, because the latter would mean a transition from a Hermitian section to NH2 from the left (in the current setup there is no such connection). 
Finally, the impedance matching of the outer ends of H1 and H2, which was not related to the nonreflecting tunneling, but was used for the sake of simulation clarity, was obtained by invoking the chain in Fig. \ref{fig:segments}(a), but with an open boundary. Then, only an incident and a reflected wave exist, leading to
\begin{equation}
    R_{edge}=\frac{i\widehat{\Omega}+\Omega^2/2-i\Omega H_v-H_y}{i\widehat{\Omega}-\Omega^2/2+i\Omega H_v+H_y} \quad \Rightarrow \quad H_v=\frac{\widehat{\Omega}}{\Omega}, \quad H_y=\frac{\Omega^2}{2}.
\end{equation}

\section{The electrical transmission-line analogue in Eq. (5)}

\begin{figure}[htpb]
\begin{center}
\setlength{\tabcolsep}{14pt}
\def\arraystretch{1.9} 
\begin{tabular}[c]{cc}
\textbf{(a)} & \textbf{(b)}   \\
   \includegraphics[height=4.6 cm, valign=c]{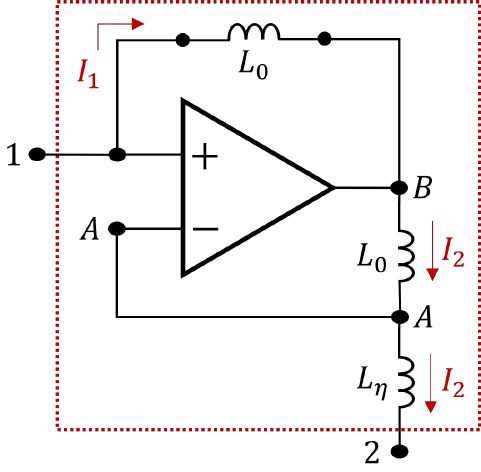}  &   \includegraphics[height=4.0 cm, valign=c]{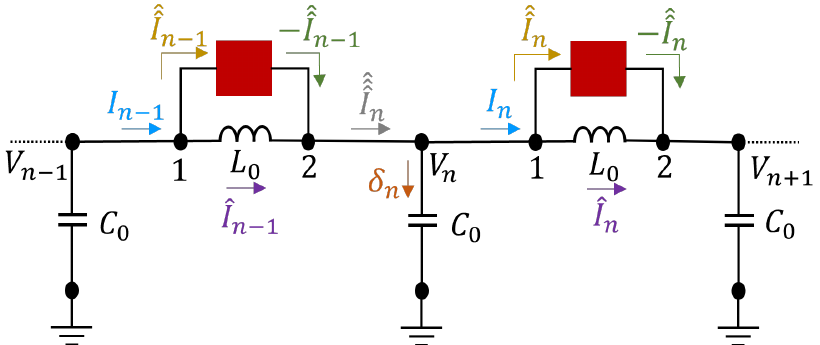} 
    \end{tabular}
\end{center}
\caption{Creating nonreciprocity in current flow using inductive negative impedance converters. (a) A standalone converter. (b) The integration of the converter in the transmission line.}
\label{fig:opamp_chain}
\end{figure}

To transform from the generic mechanical transmission line into an electrical one we substitute $M_0\leftrightarrow C_0$ and $K_0\leftrightarrow L_0^{-1}$, where $C_0$ and $L_0$ were defined in the main text as the nominal capacitor and inductor.
We then take a closer look at the operational amplifier arrangement of Fig. 3 of the main text, defining nodes $A$ and $B$ and the currents $I_1$ and $I_2$, as illustrated in Fig. \ref{fig:opamp_chain}(a) for the red cell (for brevity without the stabilizing resistors $R_p$ and $R_q$).
This arrangement works similarly to the usual negative impedance converter, just here realized with inductors rather than resistors. 
We have $\Dot{I}_1=(V_1-V_B)/L_0$ and $\Dot{I}_2=(V_B-V_A)/L_0$, which due to $V_1=V_A$ gives the expected $I_2=-I_1$. 
Since we also have $\Dot{I}_2=(V_A-V_2)/L_\eta$, we obtain
\begin{equation} \label{eq:V_B}
    V_B=V_1\left(1+\frac{L_0}{L_\eta}\right)-V_2\frac{L_0}{L_\eta}.
\end{equation}
Equating \eqref{eq:V_B} with $V_B=-L_0\Dot{I}_1+V_1$, leads to
\begin{equation} \label{eq:V_12}
    V_1-V_2=-L_\eta\Dot{I}_1.
\end{equation}
Incorporating \eqref{eq:V_12} into the transmission line, as illustrated in Fig. \ref{fig:opamp_chain}(b), the voltage gradient between the $n-1$, $n$ and the $n+1$ nodes takes the form
\begin{equation}
    V_{n-1}-V_n=-L_\eta\Dot{\widehat{\widehat{I}}}_{n-1}=L_0\Dot{\widehat{I}}_{n-1} \quad , \quad V_n-V_{n+1}=-L_\eta\Dot{\widehat{\widehat{I}}}_n=L_0\Dot{\widehat{I}}_n.
\end{equation}
Therefore, the outgoing and incoming currents of the $n_{th}$ node respectively become
\begin{subequations}  
\begin{align}    \Dot{\widehat{\widehat{\widehat{I}}}}_n&=\Dot{\widehat{I}}_{n-1}-\Dot{\widehat{\widehat{I}}}_{n-1}=(V_{n-1}-V_n)\left(\frac{1}{L_0}+\frac{1}{L_\eta}\right), \\    
\Dot{I}_n&=\Dot{\widehat{I}}_n+\Dot{\widehat{\widehat{I}}}_n=(V_n-V_{n+1})\left(\frac{1}{L_0}-\frac{1}{L_\eta}\right),
\end{align}
\end{subequations}
which for $L_\eta=\eta
^{-1}L_0$ is equivalent to Eq. (3) of the main text.
Employing then the relation $\delta_n=C_0\Dot{V}_n=\widehat{\widehat{\widehat{I}}}_n-I_n$, we obtain the voltage dynamics at the $n_{th}$ node, as
\begin{equation}
    C_0\Ddot{V}_n=(V_{n-1}-V_n)\left(\frac{1}{L_0}+\frac{1}{L_\eta}\right)-(V_n-V_{n+1})\left(\frac{1}{L_0}-\frac{1}{L_\eta}\right),
\end{equation}
which is fully equivalent to \eqref{eq:nonrecip_mechanical_red}. The equivalence for the blue chain is obtained in a similar way.


\end{document}